\documentclass[
superscriptaddress,amsmath,amssymb,aps,prd]{revtex4-1}
\usepackage{graphicx}
\usepackage{dcolumn}
\usepackage{bm}

\usepackage{acronym}
\usepackage{gensymb}
\usepackage{color}
\usepackage[colorlinks,linkcolor=blue,citecolor=blue,urlcolor=blue ]{hyperref}
\usepackage{subfigure}
\usepackage{longtable,multirow,dcolumn}
\usepackage{footnote}

\DeclareMathOperator{\sech}{sech}
\newcommand{\msun}{M$_\odot$}

\acrodef{EM}{electromagnetic}
\acrodef{DWD}{white dwarf binaries}
\acrodef{WD}{white dwarf}
\acrodef{CVBs}{candidate verification binaries}
\acrodef{GR}{general relativity}
\begin{document}

\title{Sky location of Galactic white dwarf binaries in space-based gravitational wave detection}
\author{Pan Guo}
\affiliation{The School of Physical Sciences, University of Chinese Academy of Sciences, Beijing 100049, China}
\affiliation{The International Centre for Theoretical Physics Asia-Pacific, University of Chinese Academy of Sciences, Beijing 100190, China}
\author{Hong-Bo Jin\footnote{Corresponding author Email: hbjin@bao.ac.cn}}
\affiliation{National Astronomical Observatories, Chinese Academy of Sciences, Beijing 100101, China}
\affiliation{Hangzhou Institute for Advanced Study, UCAS, Hangzhou 310024, China}
\author{Cong-Feng Qiao}
\affiliation{The School of Physical Sciences, University of Chinese Academy of Sciences, Beijing 100049, China}
\affiliation{The International Centre for Theoretical Physics Asia-Pacific, University of Chinese Academy of Sciences, Beijing 100190, China}
\author{Yue-Liang Wu}
\affiliation{The International Centre for Theoretical Physics Asia-Pacific, University of Chinese Academy of Sciences, Beijing 100190, China}
\affiliation{Hangzhou Institute for Advanced Study, UCAS, Hangzhou 310024, China}
\affiliation{CAS Key Laboratory of Theoretical Physics, Institute of Theoretical Physics, Chinese Academy of Sciences, Beijing 100190, China}

\date{\today}
\begin{abstract}
Quickly localizing the identified white dwarf (WD) binaries is the basic requirement for the space-based gravitational wave (GW) detection. 
In fact, the amplitude of GW signals are modulated by the periodic motion of GW detectors on the solar orbit. The intensity of the observed signals is enhanced according to the observation time beyond a year to enhance a high signal to noise ratio (SNR). 
As data gap exists, the completeness of the data observed for a long time depends on filling gaps in the data. 
Actually, in a year period, the GW sources have a best observation orbit position of GW detectors, where the intensity of GW is maximum.
Thus, the best positions can be searched for the verified GW sources of the sky map to enhance SNR too, which avoids filling data gaps.
For the three arms response intensity of the GW signals changing more clearly with the location of the GW sources relative to the detector, the noises and the suppression of noise by time delay interferometer are ignored.
As a verification case, the four WD binaries are chosen, whose best observation orbit positions of the GW detectors are related to the direction of WD binaries perpendicular to the detection arms. 
The two verification binaries: J0806 and V407 Vul are observed at the best orbit positions by TAIJI for the short time of 2 and 3 days respectively. The corresponding intensities of GWs are above the values of the TAIJI sensitivity curve, significantly.
Location parameter estimation of the verification WD binaries are performed using the Metropolis-Hastings MCMC method. 
The confidence level of the parameters obtained in the best position is significantly several times higher than that in the worst position where the direction of WD binaries are almost parallel to the detection arms.
Compared with a single detector, the network of two detectors improve slightly the accuracy of location of the verification binaries.  
These results imply that the searching of GW signals and parameter estimation of GW sources from the experimental data of the space-based mission do not ignore the   orbit positions relevant to GW sources.
 
\end{abstract}

\keywords{Gravitational waves, white dwarf binaries}
\pacs{95.85.Sz, 04.30.−w, 04.30.Tv} 
\maketitle

\section{Introduction}
The nearly hundred events of gravitational waves above 10 Hz are observed during the observing run O1-O3 by the ground-based observatories from LIGO, VIRGO and KAGRA collaborations \cite{LIGOScientific:2021djp,KAGRA:2023pio}. 
The gravitational wave observation below 1 Hz is a space-based mission matching the longer arms with the order of $10^6$ km apart between the spacecrafts\cite{Ruan:2020smc}, such as LISA \cite{LISA_menu2007,Audley:2017drz} and TAIJI \cite{10.1093/nsr/nwx116}.
TianQin\cite{TianQin:2015yph,TianQin:2020hid} is also space-based gravitational wave observatory, which has $10^5$ km triangular arms on high Earth orbit. 
The frequency band observable by DECIDO \cite{Sato:2009zzb, Kawamura:2020pcg}, BBO \cite{Cutler:2009qv}, TianGO \cite{Kuns:2019upi} is around 0.1 Hz for observing binary white dwarf mergers through gravitational waves\cite{Kinugawa:2019uey}.

The gravitational wave sources detected by the ground-based observatories and the space-based mission have the different responses to the detection arms. 
The cross-correlated response of the Hanford and Livingston LIGO detectors is modulated as the rotation of the Earth sweeps the antenna pattern across the sky\cite{Allen:1996gp,Cornish:2001hg}.
Referring to the configuration of LISA \cite{LISA_menu2007,Audley:2017drz} and TAIJI \cite{10.1093/nsr/nwx116}, the space-based observatories consist of a triangular structure of three spacecrafts in a heliocentric orbit with an arm length of the million kilo-meters. 
The three spacecrafts orbit Kepler, with the same annual period as Earth, have two forms of motion, one is the revolution of the heliocentric orbit, and the other is the rotation of three spacecraft around the center of the triangle\cite{Ruan:2020smc}.
The relative velocity between the two spacecrafts causes Doppler shift of the photon emitted by the first spacecraft and received by the second. The heterodyne interferometry is required to track the relative displacement between two spacecrafts. 
The on-orbit detectors have maximal response of the gravitational wave as a transverse wave, when the propagation orientation of the gravitational wave source is perpendicular to the sides of a triangle made up of three spacecrafts. On the contrary, if it is parallel, there is almost no response. Thus, The amplitude of gravitational waves is modulated with a year period.

The orbital modulation effect of the observed gravitational wave signals is described by the detector response function. 
The full LISA response function to an arbitrary gravitational wave is derived using a coordinate free approach in the transverse-traceless gauge\cite{Cornish:2002rt}. The forward modeling of space-based gravitational wave detectors was proposed and an adiabatic approximation to the detector response significantly extends the range of the standard low frequency approximation\cite{Rubbo:2003ap}. 
Recently, a time-domain generic response code LISA gravitational wave Response\citep{LISAGWResponse_2022_6423436} is proposed for projecting the gravitational wave polarizations onto the LISA constellation arms.
In this paper, the TAIJI response function is expressed in the transverse-traceless gauge. The length of on-orbit detection arms in space-time is calculated numerically in the time domain. 

The GW signals from Massive Black Hole merging linger for much longer in the detector sensitive band, which overlap with the continuous and nearly monochromatic GW signals of WD binaries. The data analysis of those overlapping signals is one of the Mock LISA Data Challenges\cite{Babak:2009cj}. These overlapped signals cross the entire TAIJI frequency spectrum too. In fact, for the third-generation ground-based detectors the signals from different sources will be simultaneously present in the data, whose associated computational challenges have recently begun\cite{Speri:2022kaq}. The data analysis of overlapped GW signals become already a general issue to be solved. The TAIJI Data Challenges (TDC) are online for the solution of that issue\cite{Ren:2023yec}. Considering the location parameters of the detectors, the overlapped GW signals may be degenerated slightly for a verification source.

Among all kinds of ultra-compact stellar mass binaries, \acp{DWD} comprise the absolute majority (up to $10^8$) in the Milky Way. Being abundant and nearby, \acp{DWD} are expected to be the most numerous gravitational wave sources for space-based detectors \cite{Nelemans:2001b,Yu:2010,Breivik:2019,Lamberts:2018}.
There are many predictions based on a Galaxy model combined with a binary population model~\citep{Nelemans:2001a,Nelemans:2004,Nissanke:2012,Ruiter:2010,Postnov:2014,Toonen:2017}. Recently, more studies construct an observationally driven population \citep{Korol17_WDLISA,Breivik:2019,Korol_19_disk_bulge,Korol:2021pun}. The numerous monochromatic gravitational waves emitted by the Galactic \acp{DWD} form the foreground signals of sky map. In these sources, the verification binaries (VBs) signals, such as J0806, V407 Vul, ES Cet and SDSSJ1351(abbr of SDSSJ135154.46-064309.0) have been discovered by X-ray and optical observatories \cite{Kupfer:2018jee}. These binary sources have the rotational period ranging from a few hundred to a thousand seconds, that are sensitive to the detection arms of LISA and TAIJI. 

Quickly localizing the identified binaries is the basic requirement for the space-based gravitational wave detection, that is used to calibrate the detection data and reduce the noises of detectors. Furthermore, through the localizing the sky position of the gravitational wave sources, on the one hand, it is possible to cross-check the verification binaries. On the other hand, more new unbiased samples will be detected to supplement the astronomical observation results of verification binaries.
As a Galactic population, the density of VBs reaches a peak near the Galactic Plane. However, most the known systems are located in the Northern Hemisphere, with only a few systems located at low Galactic latitudes, indicating that the current sample may be very incomplete and biased\cite{Kupfer:2018jee,Kupfer:2023nqx}. The optical surveys such as OmegaWhite \cite{2016MNRAS.463.1099T}, ZTF \cite{2019PASP..131a8002B}, LAMOST\cite{Zhao:2012nm}, SDSS-V\cite{2019BAAS...51g.274K}, etc are expected to find a large sample of multi-messenger sources \cite{Korol17_WDLISA,Li:2020voo} to supplement the astronomical observation results of verification binaries as a large sample of multi-messenger sources. 

As three spacecrafts are moving on the solar orbit with a year period, the orientation of the identified binaries does not keep being perpendicular to the detection arms. The detection arm response intensity of the gravitational wave of the observed sources have the maximum and minimum relative to the perpendicular and parallel to detection arms. Thus, a gravitational wave source have a best observed window in a year period of orbit. 

Generally, on the source location of gravitational wave signal, the intensity of the observed signal is enhanced according to the observation time (integration time) to obtain a high SNR. The observation time is generally selected as years, so the modulation effect of the detector at different orbital positions of the gravitational wave source is ignored.As the data gap exists in the whole year time, a long time observation data need filling the gap, which takes deviation from the real data\cite{Speri:2022kaq}. In this paper, we consider the modulation effect of the detector, i.e., the orbit position of space-based laser interferometer detectors, have significant effects on the the localizing the sky position of the identified binaries.
For the three arms response intensity of the GW signals changing more clearly with the location of the GW sources relative to the detector, the noises and the suppression of noise by time delay interferometer are ignored.
We focus on the projected signal on the single arm detector of an identified WD binary. So, instead of a few years, observation time  was replaced with a short one.  

Localizing the sky position of the gravitational wave source is a key scientific goal for gravitational wave observations\citep{Blaut:2011zz}.
There are two major methods for localizing the sky position of GW sources. One method is the Fisher information matrix approximation (FIM), which can give robust estimation of the sky localization of the source with a high SNR where the inverse of FIM gives the covariance matrix of the parameters.
Another method is the Bayesian estimation i.e. the posterior distribution expressing the uncertainty of the source parameters\citep{Shuman2022}.
The angular resolution is often chosen as the sky localization accuracy for an arbitrary network of interferometric gravitational-wave (GW) detectors\citep{Wen2010}.
Angular resolution of the detector and the estimation errors of the signal’s parameters in the high frequency
regimes are calculated as functions of the position in the sky and as functions of the frequency\citep{Zhang:2020hyx}. In addition, detectors configuration properties, such as the orientation change of the detector plane, can also effect the angular resolution\citep{Zhang:2020hyx}. Besides, the network for different two detectors can effectively help to get better accuracy in the sky localization. As an example,
similarities and complements between LISA and TAIJI imply that LISA-TAIJI network can effectively help to accurately localize GW sources,
since the angular resolution measurements for the network depend on the configuration angle and separation of the two constellations\citep{Ruan_2021,Ruan2020}. The network was estimated to improve the angular resolution for over 10 times  by comparison with each individual LISA or TAIJI detector\citep{Cutler:1997ta}.
The LISA-TianQin network has better ability in sky localization for sources with frequencies in the range 1-100 mHz and the network has larger sky coverage for the angular resolution than the individual detector\citep{Wang:2020vkg}. 

For the first time, we focus on the sky location of the identified binary sources in a short observation time, considering the amplitude modulation effect of different detector positions on the WD binary response.
Due to the periodic orbit motion of the detector, the results of the short time observation are related to the orbit position of the detector. For some specific WD binary sources, the best localizing the sky position of the sources at the best detector position is achieved in this paper.
As a verification case, the four WD binaries are chosen, whose best observation orbit positions of the GW detectors are related to the direction of WD binaries perpendicular to the detection arms. 
The two verification binaries: J0806 and V407 Vul are observed at the best orbit positions by TAIJI for the short time of 2 and 3 days respectively, whose GW intensities are greater than the values of TAIJI sensitivity curve, significantly. The other two verification binaries: Es Cet and SDSSJ1351 are beyond the sensitive cure for the observation time of 35 and 52 days respectively when the their intensities are above TAIJI sensitivity curve. 
Compared with a single detector, the network of two detectors improve slightly the accuracy of location of the verification binaries. The reason of that result is that one GW source can not be perpendicular to both detectors of TAIJI and LISA for a short observation time. For a long observation time, the network of two detectors has a significant improvement to angular resolution.   

The structure of this paper is as follows: in Sec.\ref{Sdwdb} , we mainly introduce the signal model of the WD binary star and the distribution of its constituent parameters; in Sec.\ref{Sdr} , we mainly introduce the detector orbit and the detector response form in the time domain; in Sec.\ref{sec:lsp} , the best localizing the sky position of the sources at the best detector position is achieved; in Sec.\ref{sec:conclusion} , the full text is summarized. 

\section{Detector response } \label{Sdr}

\subsection{Detector Orbit} \label{Sdr_oe}
Referring to LISA orbit, the TAIJI spacecrafts will orbit Kepler\cite{Rubbo:2003ap} and the arm length is 3$\times 10^6$ km, which is different from LISA arm length: 2.5 $\times 10^6$ km. Each spacecraft positions are expressed as a function of time. To second order in the eccentricity, the Cartesian coordinates of the spacecraft are given by
\begin{eqnarray} \label{keporb}
    x(t) &=& R \cos(\alpha) + \frac{1}{2} e R\Big( \cos(2\alpha-\beta) -
    3\cos(\beta) \Big) \nonumber\\
    && + \frac{1}{8} e^2 R\Big( 3\cos(3\alpha-2\beta) - 10\cos(\alpha)
    \nonumber\\
    && - 5\cos(\alpha-2\beta) \Big) \nonumber\\
    y(t) &=& R \sin(\alpha) + \frac{1}{2} e R\Big( \sin(2\alpha-\beta) -
    3\sin(\beta) \Big) \nonumber\\
    && + \frac{1}{8} e^2 R\Big( 3\sin(3\alpha-2\beta) - 10\sin(\alpha)
     \nonumber\\
    && + 5\sin(\alpha-2\beta) \Big) \nonumber\\
    z(t) &=& -\sqrt{3} e R\cos(\alpha-\beta) \nonumber\\
    &&+ \sqrt{3} e^2 R \Big( \cos^2(\alpha-\beta)
    + 2\sin^2(\alpha-\beta) \Big) 
\end{eqnarray}
Where $R = 1$ AU is the radial distance to the guiding center, $e$ is the eccentricity, $\alpha = 2\pi t/1year + \kappa$ is the orbital phase of the guiding center, and $\beta = 2\pi n/3 + \lambda$ ($n=0,1,2$) is the relative phase of the spacecraft within the constellation.  The parameters $\kappa$ and $\lambda=0$ give the initial ecliptic longitude and orientation of the constellation. The orbital eccentricity is computed based on the arm length: $e = L/(2 \sqrt{3} R )$.  
By setting the mean arm-length equal to those of the TAIJI baseline, $L = 3 \times 10^9$ m, the spacecraft orbits are found to have an eccentricity of
$e=0.005789$, which indicates that the second order effects are down by a factor of 100 relative to leading order.

TianQin is composed of three drag-free spacecrafts in an equilateral triangular constellation orbiting around the Earth. The guiding center of the constellation coincides with the geocenter. The geocentric distance of each spacecraft is $1.0 \times 10^5$ km, which makes the distance between each pair of spacecrafts be  $1.7 \times 10^5$ km. The period of the nearly circular Keplerian orbit of the spacecraft around the Earth is approximately 3.65 day. More details are found in the paper\cite{Hu:2018yqb}.
The constellation plane of TianQin is nearly vertical to the ecliptic and faces the white-dwarf binary RX J0806.3+1527 (abbreviated with J0806) as a reference source\cite{Ye:2020tze}. Thus, the GW amplitude of J0806 is not modulated by the different detector orbit positions. As the orbit period of TIANQIN is 3.65 day\cite{Hu:2018yqb}, the other GW sources has the different modulation effect from TAIJI and LISA.    

\subsection{Response Function} \label{Sdr_rf}
The gravitational waves are expressed as the perturbations of the flat space-time
\begin{equation}
    g_{\mu \nu} =\eta_{\mu \nu}+h_{\mu \nu}
\end{equation}
The laser link signal is emitted from the spacecraft $i$ at $t_i$ moment and received by the spacecraft $j$ at $t_j$ moment. Assume that $c=1$. Then, the distance $\ell_{i j}$ between the the laser link in space-time can be obtained~\cite{Rubbo:2003ap}

\begin{eqnarray}
    \ell_{i j} = \int_{i}^{j}  \sqrt{g_{\mu \nu} d x^\mu d x^\nu} = \left | {\mathbf{x}_j(t_j)-\mathbf{x}_i(t_i)}  \right | \nonumber\\
    + \frac{1}{2} \hat{r}_{i j}(t) \otimes \hat{r}_{i j}(t) : \int_{i}^{j} {\bf h}(\xi (\lambda)) d\lambda
\end{eqnarray}
Here $\left | \mathbf x-\mathbf y \right |$ denotes  the  Cartesian  distance between $\mathbf x$ and $\mathbf y$ and ${\bf h}(\xi)$ is the gravitational wave tensor in the transverse-traceless gauge.  The colon here denotes a double contraction, ${\bf a}:{\bf b} = a^{i j}b_{i j}$ .

$\hat{r}_{i j}(t)$ denotes the unit vector
\begin{equation}
    \hat{r}_{i j}(t_i) = \frac{{\bf x}_j(t_j) - {\bf
        x}_i(t_i)}{\ell_{i j}} \, .
\end{equation}

So the equation is used to calculate $\delta \ell_{i j}$
\begin{equation} \label{deltaL1}
    \delta \ell_{i j} =\frac{1}{2} \hat{r}_{i j}(t) \otimes \hat{r}_{i j}(t) : \int_{i}^{j} {\bf h}(\xi (\lambda)) d\lambda
\end{equation}
When the gravitational wave traveling  in  the $ \hat k$ direction, the  wave  variable $\xi (\lambda)$ can be calculated by
\begin{equation}
    \xi (\lambda) = t(\lambda)-\hat{k} \cdot {\mathbf x (\lambda)}
\end{equation}
Explicitly, the time and position depend on the parameterization in the following way
\begin{eqnarray}
    t (\lambda) = t_i + \lambda \nonumber\\
    \mathbf x (\lambda) = \mathbf x (t_i) + \lambda \hat{r}_{i j} (t_i) .
\end{eqnarray}

The time and position are gotten by $d \lambda=d t$:
\begin{eqnarray}\label{xt}
    \mathbf x (t) = \mathbf x (t_i) + (t - t_i) \hat{r}_{i j} (t_i)\nonumber\\
    \xi (t) = t-\hat{k} \cdot {\mathbf x (t)}.
\end{eqnarray}
So the Eq.~(\ref{deltaL1}) can be changed to the following form
\begin{equation} \label{deltaL2}
    \delta \ell_{i j} =\frac{1}{2} \hat{r}_{i j}(t) \otimes \hat{r}_{i j}(t) : \int_{t_i}^{t_j} {\bf h}(t-\hat{k} \cdot {\mathbf x (t)}) d t
\end{equation}
The Eq.~(\ref{deltaL2}) can calculate the $\delta \ell_{i j}$ in a numerical way in the time domain.

\subsection{General analysis of annual modulation effect on detectors}\label{Sds_ga}
\begin{figure*}
    \centering
    \vspace{-0.35cm}
    \subfigtopskip=2pt
    \subfigbottomskip=2pt
    \subfigcapskip=-5pt
    \subfigure[$\kappa=0^{\circ}$]{
    \includegraphics[width=0.45\textwidth]{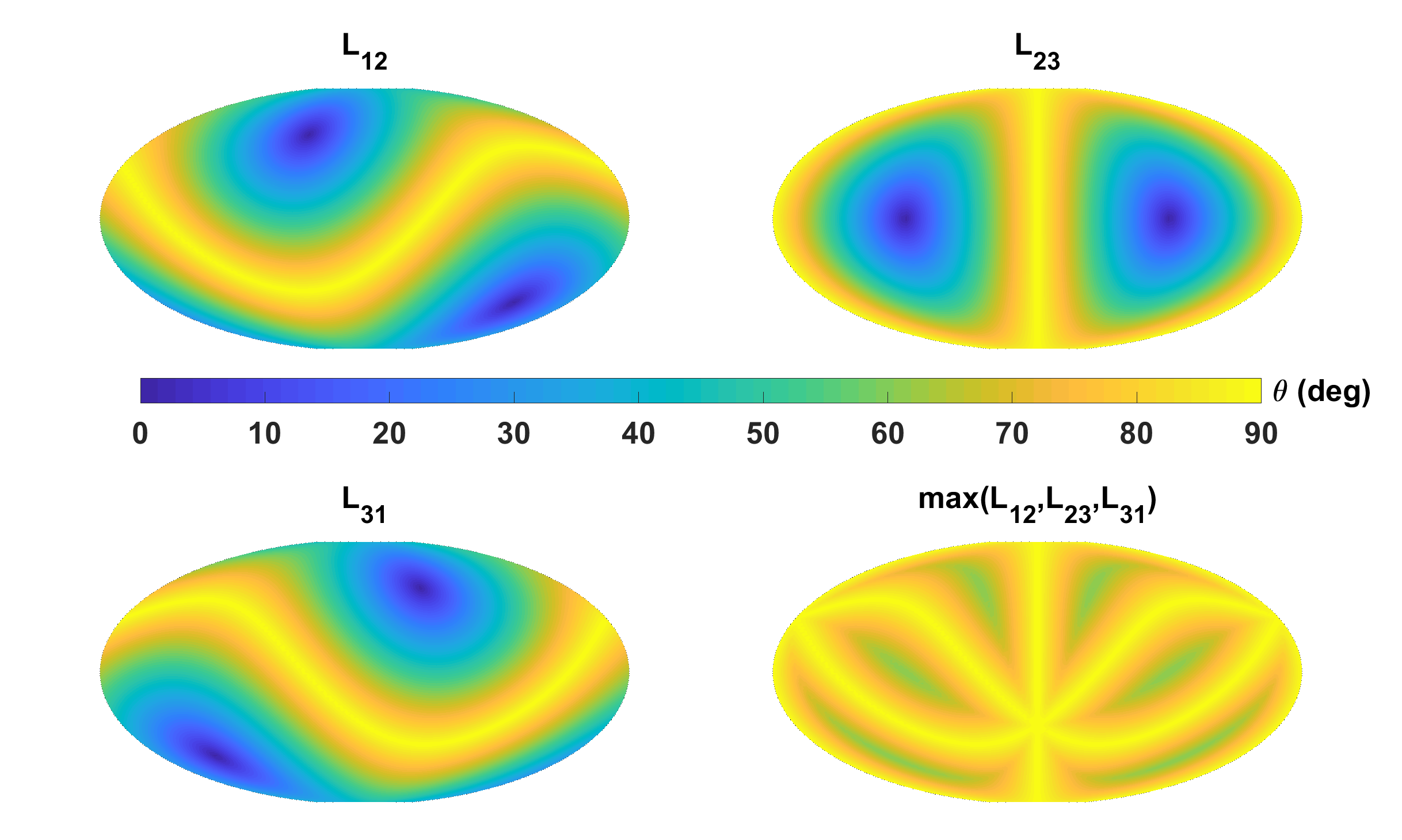}
    }
    \quad
    \subfigure[$\kappa=45^{\circ}$]{
    \includegraphics[width=0.45\textwidth]{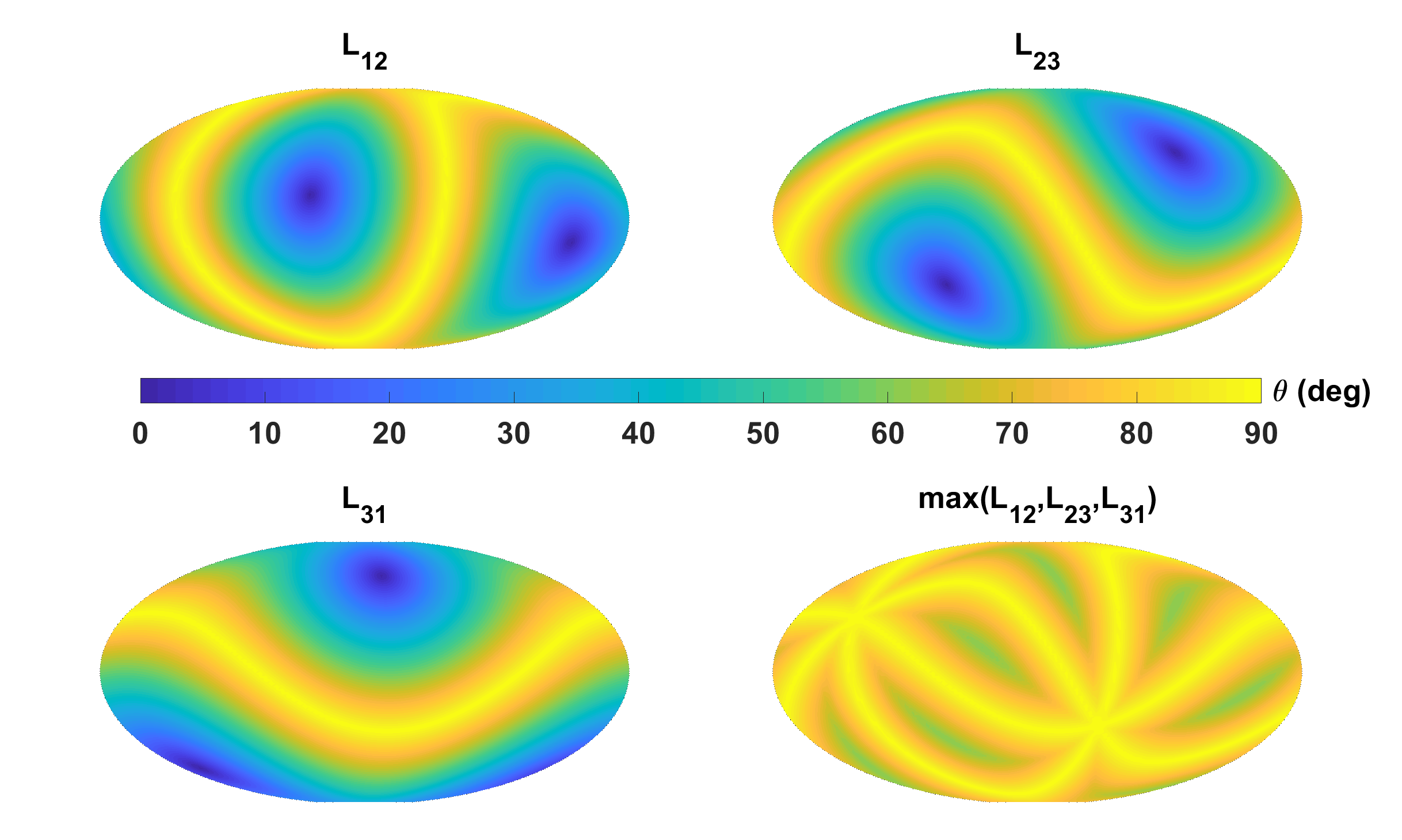}
    }
   
    \subfigure[$\kappa=90^{\circ}$]{
    \includegraphics[width=0.45\textwidth]{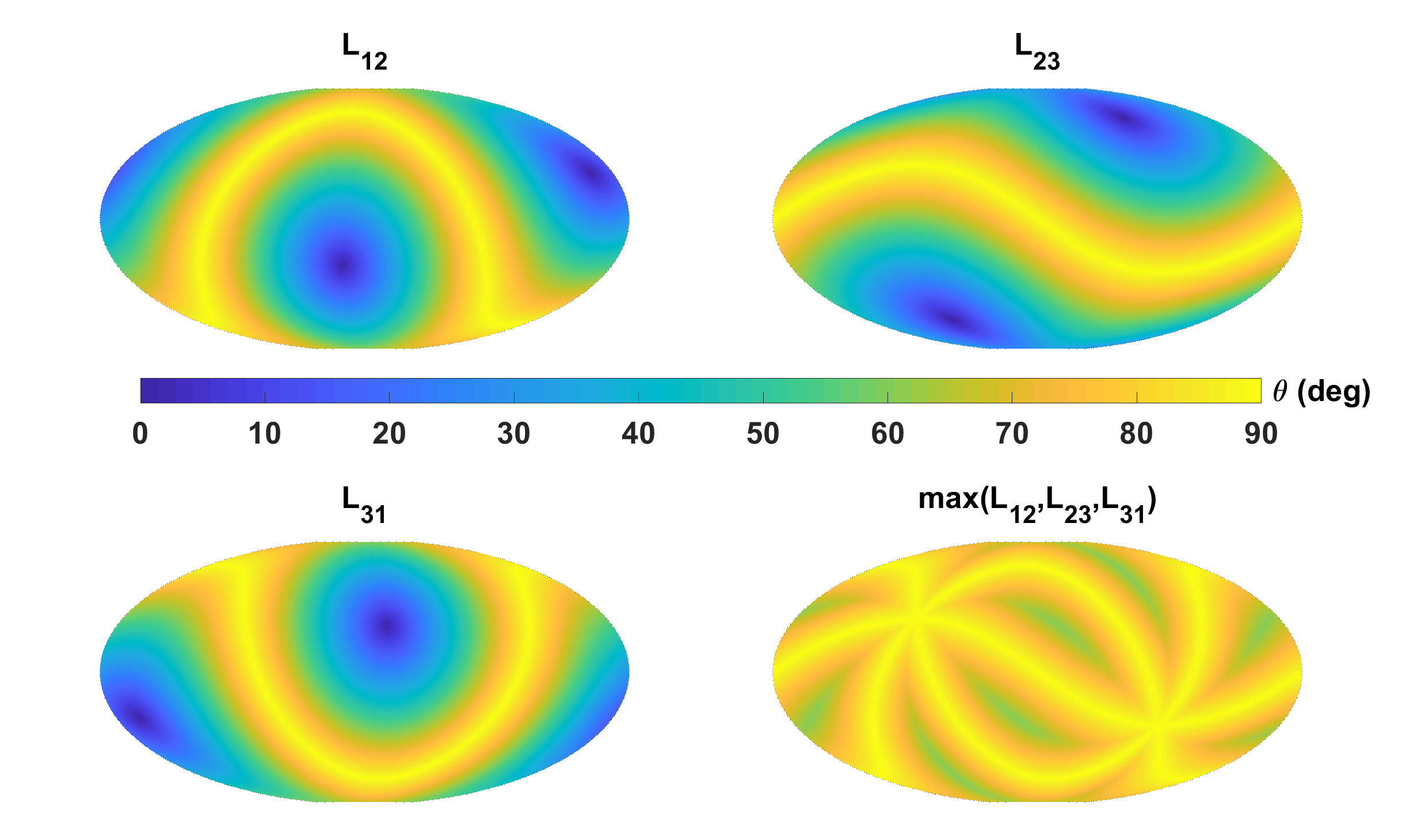}
    }
     \quad
    \subfigure[$\kappa=135^{\circ}$]{
    \includegraphics[width=0.45\textwidth]{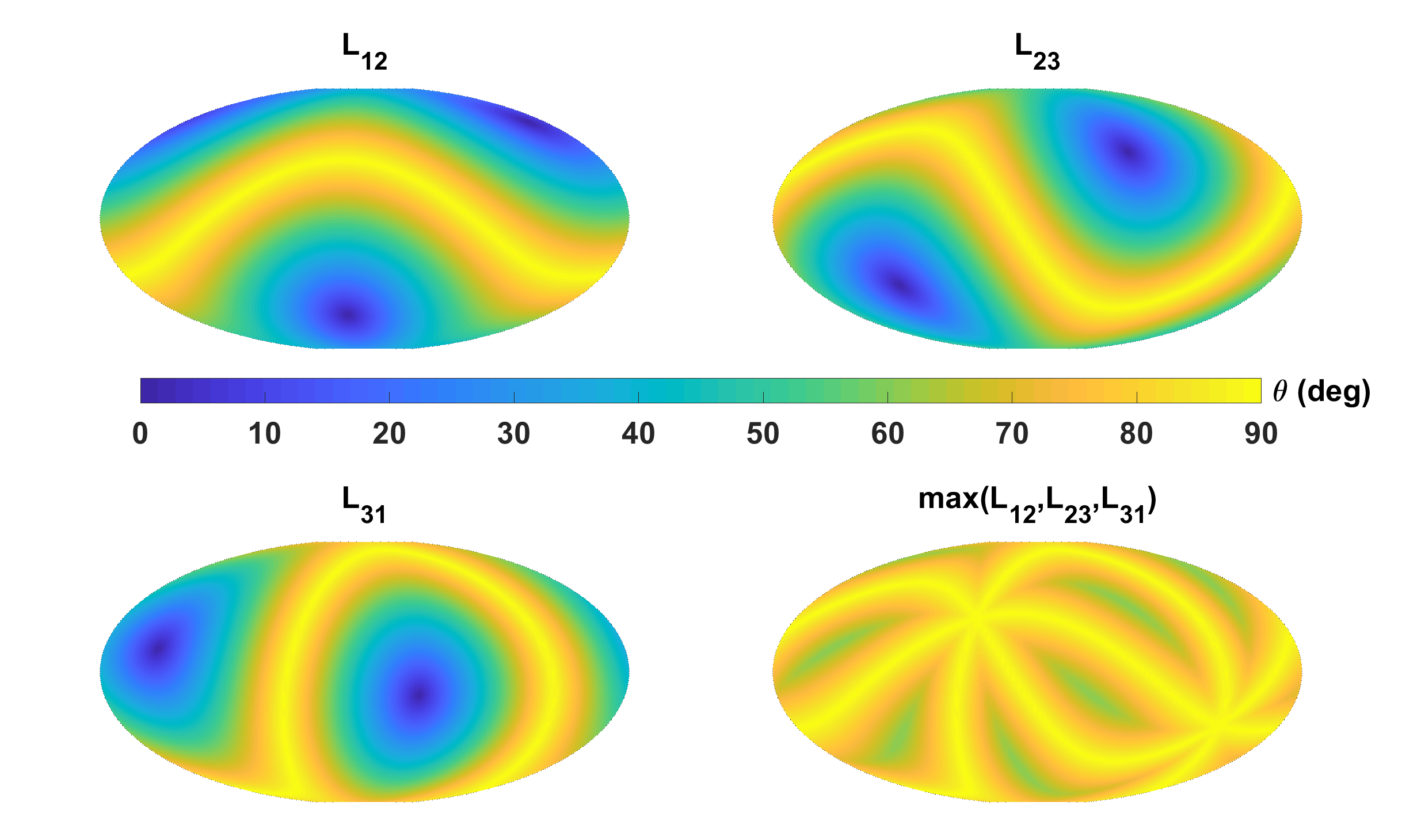}
    }
     \caption{The sky map of the angle $\theta$ between GW sources in the whole space and three detection arms ($L_{12}$, $L_{23}$, and $L_{31}$) and the maximum angle $\theta$ in three detection arms ($L_{12}$, $L_{23}$, and $L_{31}$) for different variable $\kappa$ respectively. The variable $\kappa$ ($\kappa \in \left[ 0,2\pi\right]$) is divided into 8 equal parts. In the figure, there are four values: $0^{\circ}, 45^{\circ}, 90^{\circ}, 135^{\circ}$. }
     \label{fig:skymap_Lij_1}
\end{figure*}    

\begin{figure*}
    \centering
    \vspace{-0.35cm}
    \subfigtopskip=2pt
    \subfigbottomskip=2pt
    \subfigcapskip=-5pt
    \subfigure[$\kappa=180^{\circ}$]{
    \includegraphics[width=0.45\textwidth]{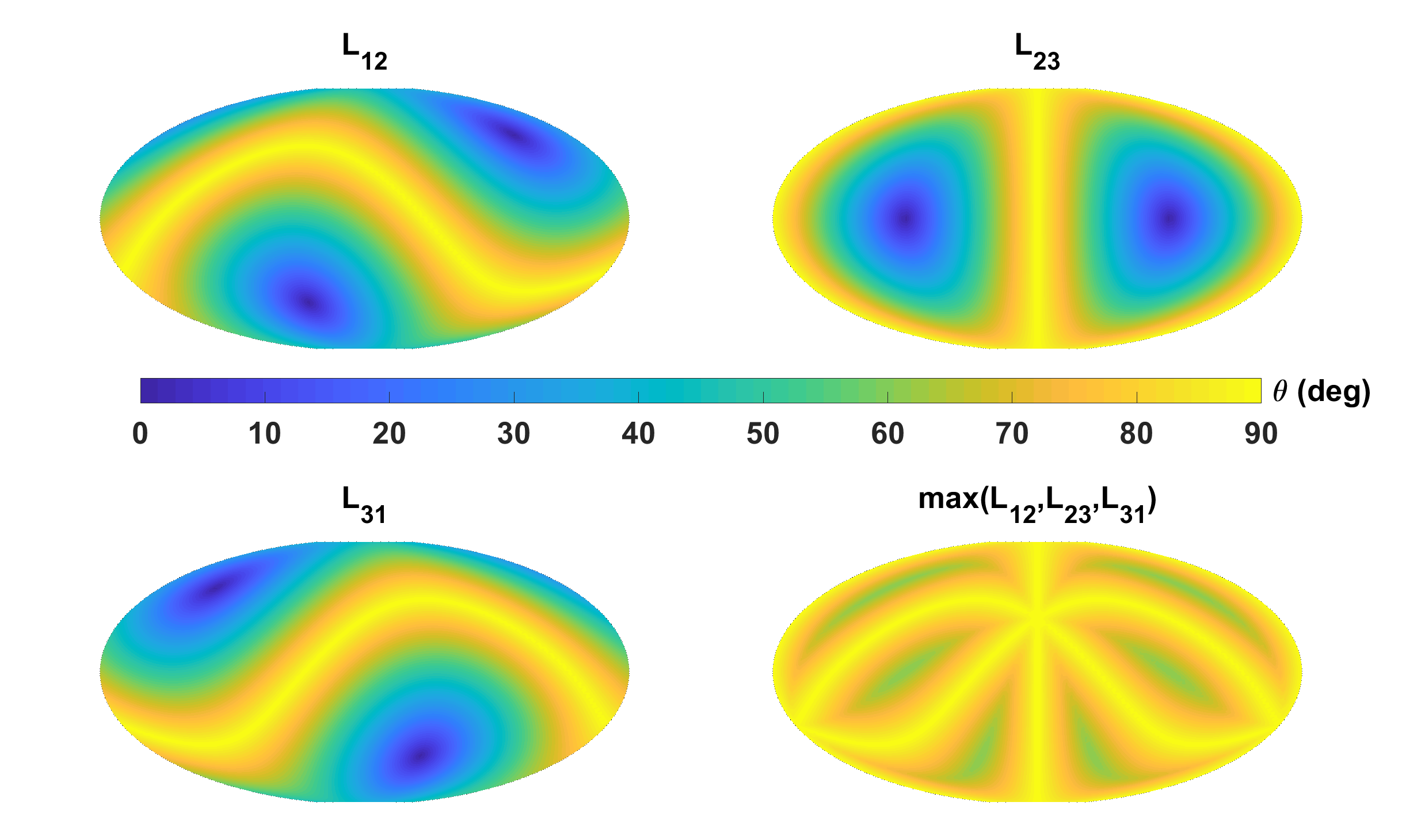}
    }
    \quad
    \subfigure[$\kappa=225^{\circ}$]{
    \includegraphics[width=0.45\textwidth]{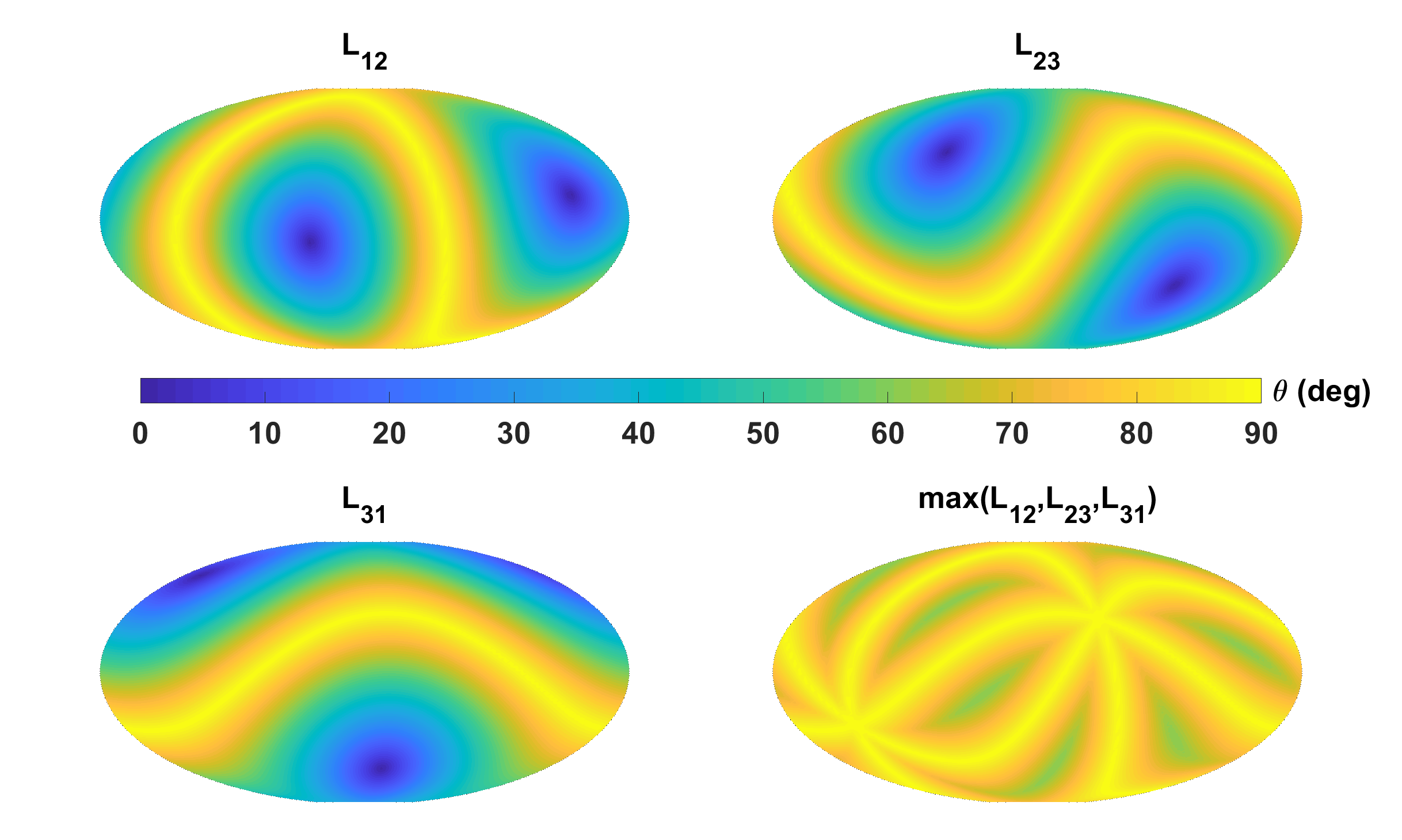}
    }
   
    \subfigure[$\kappa=270^{\circ}$]{
    \includegraphics[width=0.45\textwidth]{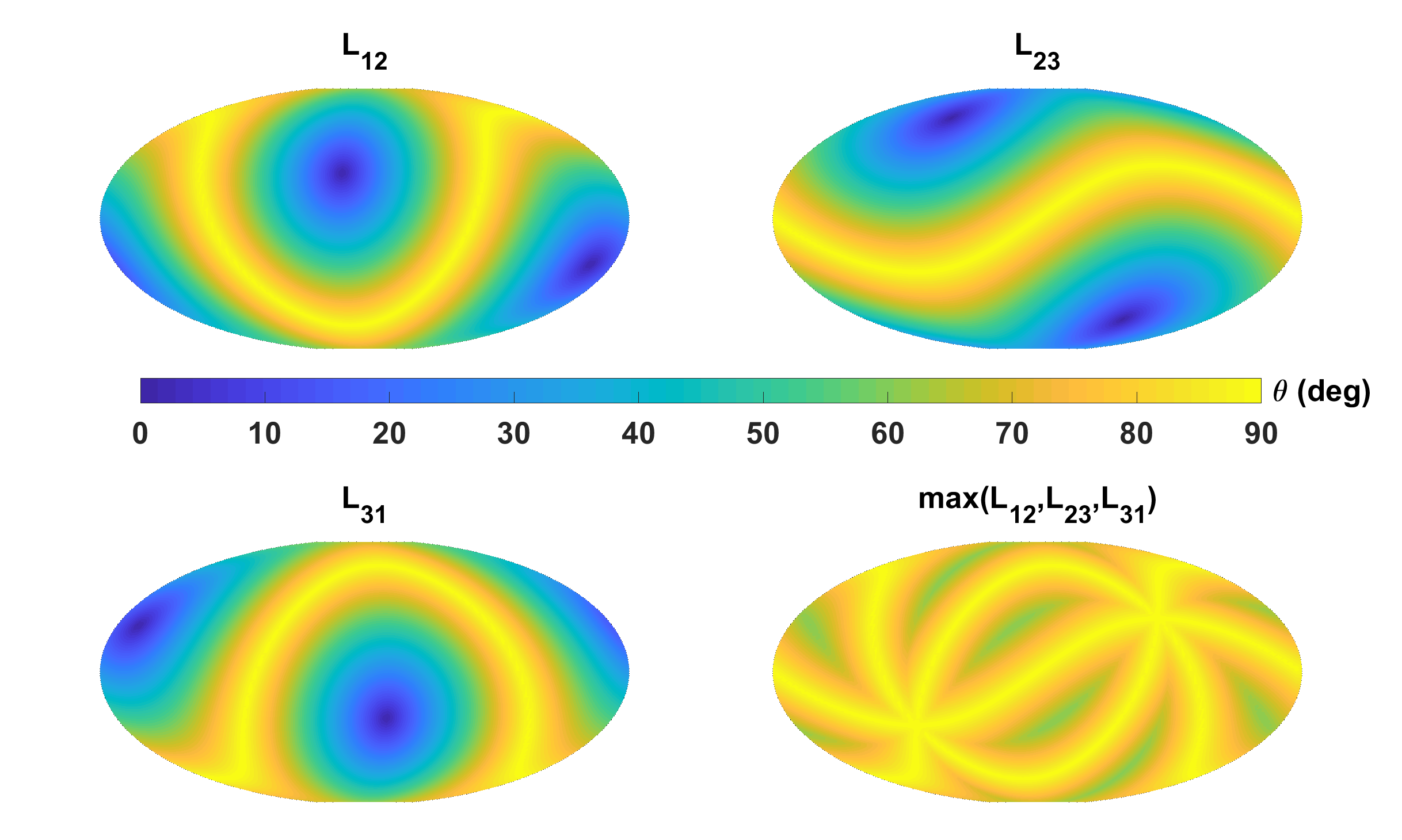}
    }
     \quad
    \subfigure[$\kappa=315^{\circ}$]{
    \includegraphics[width=0.45\textwidth]{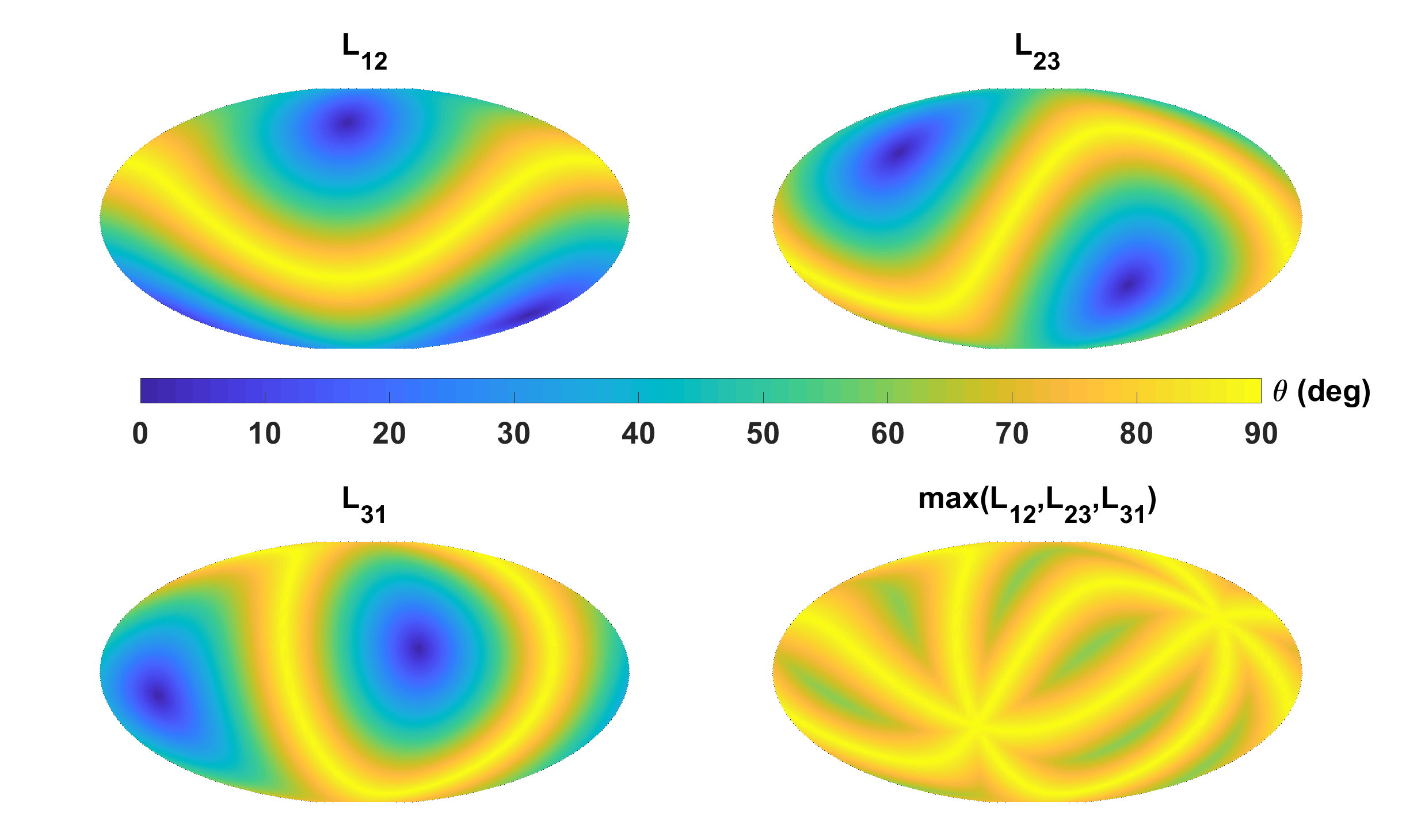}
    }
     \caption{The same as Fig.~\ref{fig:skymap_Lij_1}. There are the rest four values: $180^{\circ}, 225^{\circ}, 270^{\circ}, 315^{\circ}$.}
     \label{fig:skymap_Lij_2}
\end{figure*}

One year of period modulation effect is mainly relevant to the change of the angle $\theta$ between the propagation direction of gravitational wave source and the detection arm during the one-year orbit. The angle $\theta$ is gotten by the expression
\begin{equation}
    \theta = \arccos{\left |\frac{\hat{k}\cdot\hat{r}_{ij}}{ | \hat{k}  |\left |\hat{r}_{ij} \right | }\right |}, 
\end{equation}
where $\hat{k}$ is the propagation direction of the GW source and $\hat{r}_{i j}(t)$ denotes the unit vector of the detection arm.
\begin{equation}
    \hat{r}_{i j}(t_i) = \frac{{\bf x}_j(t_j) - {\bf
        x}_i(t_i)}{\ell_{i j}} \, .
\end{equation}
Here, we set $t_i=0$ and $t_j=\ell_{i j}/C$ (C is light velocity). The angle $\theta$ is different when the detectors are at different orbit positions. In the extreme case, when the angle $\theta=0^{\circ}$, the response of the gravitational wave to the detection arms ($L_{12}$, $L_{23}$, and $L_{31}$) is the weakest; when the angle $\theta=90^{\circ}$, that response is strongest. In this case, the ecliptic longitude $\kappa$ is used to present the detector position on orbit. 

The coordinates under the $SSB$ frame are composed of the spatial position($\beta,\lambda$). In every sky map, the horizontal axis represents the longitude ($\lambda \in \left[ - \pi, \pi\right]$)  and the vertical axis represents the ecliptic latitude ($\beta\in \left[ -\pi/2,\pi/2\right]$). 
In Fig.~\ref{fig:skymap_Lij_1} and  Fig.~\ref{fig:skymap_Lij_2}, eight sky maps are drawn for each detection arm ($L_{12}$, $L_{23}$, and $L_{31}$), which is relevant to the variable $\kappa$ ($\kappa \in \left[ 0,2\pi\right]$). In Fig.~\ref{fig:skymap_Lij_1} and  Fig.~\ref{fig:skymap_Lij_2}, the single arm response only cover the parts of all sky map, which look like the twisted long bands relevant to an area greater than 60$^\circ$. The response intensity of gravitational wave source in the area less than 60$^\circ$ (cos\ 60$^\circ$=1/2) is reduced by more than half. 
By combining the three arms, there are most of the sky map where the response intensity of the gravitational wave source is barely attenuated. In the sky map, there are also the overlapped parts where each response of the three arms is almost same intensity. 
These overlapped parts change with the the detector position expressed by the ecliptic longitude $\kappa$. 
In Fig.~\ref{fig:skymap_Lij_1} and Fig.~\ref{fig:skymap_Lij_2}, there are one or two overlapped area when the detector position changes on orbit, 
which is relevant to the selected $\kappa$: $0^{\circ},\ 45^{\circ},\ 90^{\circ},\ 135^{\circ},\ 180^{\circ},\ 225^{\circ},\ 270^{\circ}$ and $315^{\circ}$. 
In the overlapped area the gravitational wave sources have maximal response to the detector arms. That means that there are the best on-orbit positrons, which is most sensitive to the gravitational wave sources.

\begin{figure*}
\includegraphics[width=0.89\textwidth]{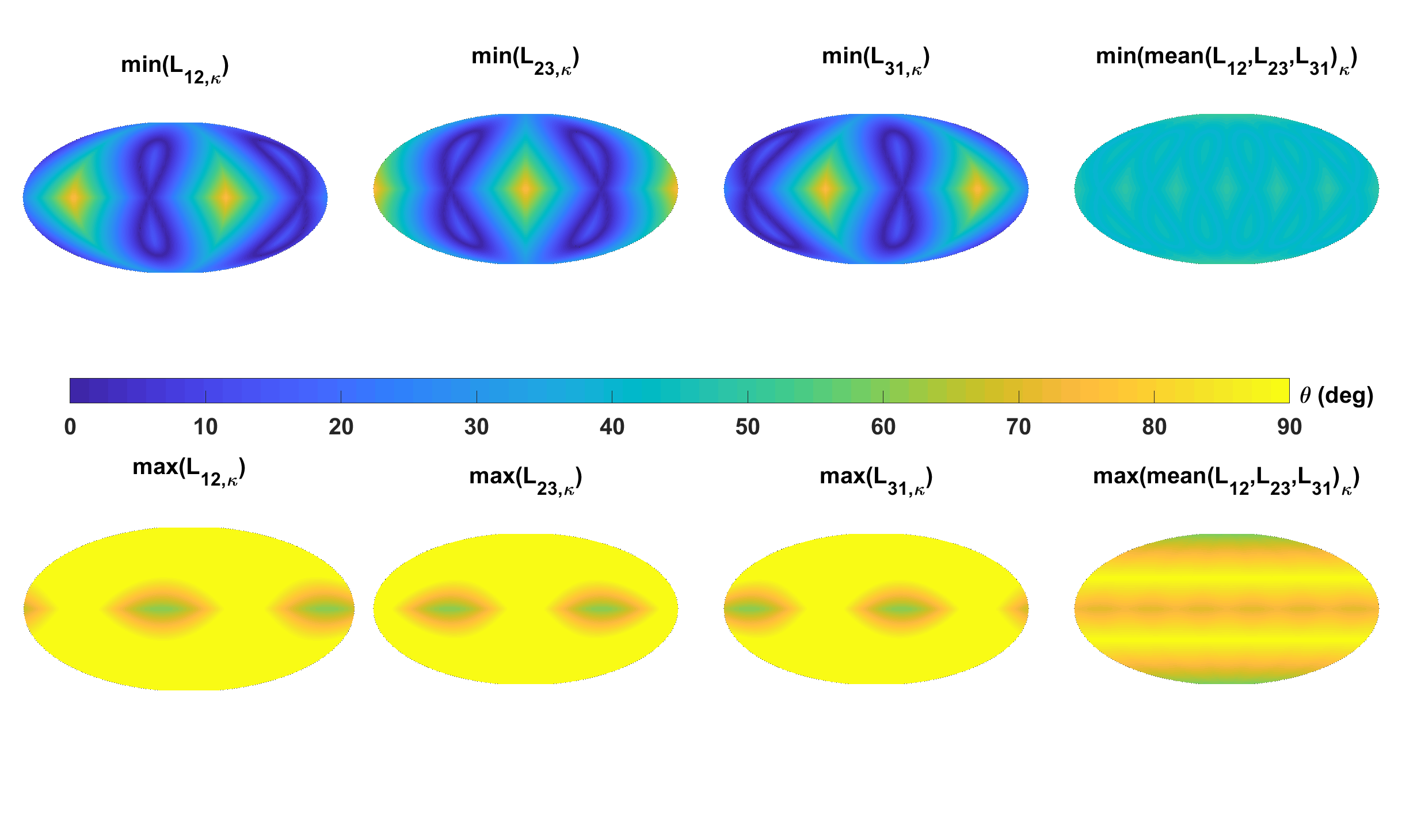}
\caption{The sky maps of the angle $\theta$ for all $\kappa$. The top three ones are the minimum-scanning by $\kappa$ value varying, which is relevant to $L_{12,\kappa}$, $L_{23,\kappa}$ and $L_{31,\kappa}$, respectively. The top right one is the combination result with three arms. The bottom t three ones are the minimum-scanning by $\kappa$ value varying, which is relevant to $L_{12,\kappa}$, $L_{23,\kappa}$ and $L_{31,\kappa}$, respectively.The bottom right one is also the combination result.}
 \label{fig:skymap_confusion}
\end{figure*}

In order to search the overlapped area in all sky map, in all orbit positions, the minimum and maximum responses, expressed with $\theta$, of the three arms are drawn in Fig.~\ref{fig:skymap_confusion}. As is seen in Fig.~\ref{fig:skymap_confusion} that there are the orbit positions relevant to the minimum responses where the the gravitational wave source's observation is difficult. Otherwise, there are also the orbit positions relevant to the maximum responses where the the gravitational wave source may be observed easily. In detail, there are some small sky where the $\theta$ are in the range of $70^{\circ}$ to $80^{\circ}$ .
It implies that the overlapped area almost covers all sky map. Thus, The best observation position on orbit, expressed with $\kappa$, is always found for a particular gravitational wave source. 

\begin{figure}
\includegraphics[width=0.89\textwidth]{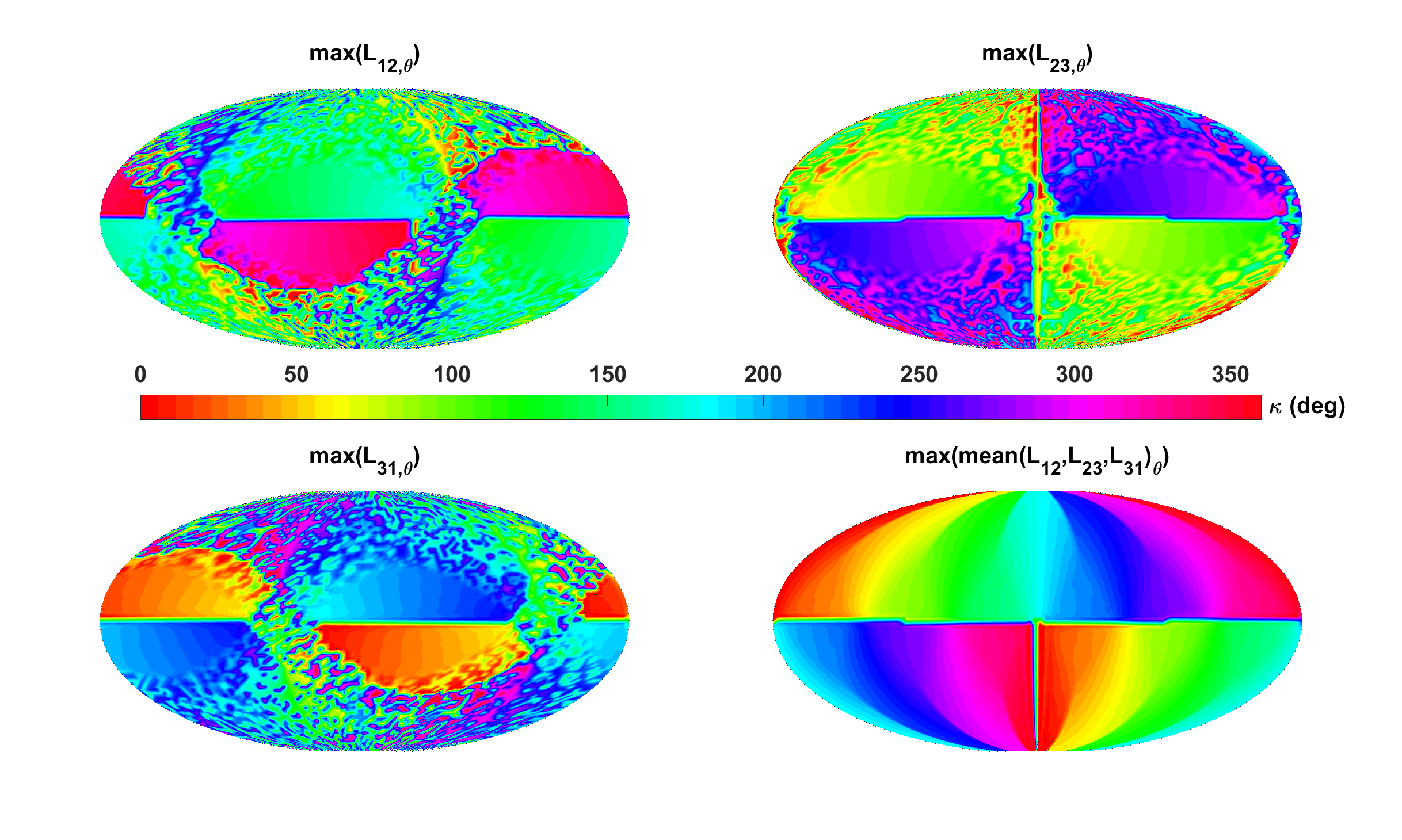}
\caption{The sky maps of $\kappa$ at the maxmum of the Angle $\theta$ of $L_{12,\kappa}$, $L_{23,\kappa}$,  $L_{31,\kappa}$, and the mean of the Angle $\theta$ of $L_{12,\kappa}$, $L_{23,\kappa}$ and $L_{31,\kappa}$ respectively. }
 \label{fig:skymap_kappa_thetaMax}
\end{figure}

\section{GALACTIC DOUBLE WHITE DWARF BINARIES}\label{Sdwdb}
\subsection{WD binary signal model}\label{Ssm_dsm}
The gravitational wave signals from the WD binary are described by a set of eight parameters: frequency $f$, frequency derivative  $\dot{f}$, amplitude ${\cal A}$, sky position in ecliptic coordinates $(\lambda,\beta)$, orbital inclination $\iota$, polarisation angle $\psi$, and initial orbital phase $\phi_0$ \citep{Cutler:1997ta,roe20,kar21}. The recipes for generating ${\cal A}$, $f$, $\dot{f}$, and $(\lambda,\beta)$ are based on the currently available observations. 
The cosine of the inclination,$\cos{\iota}$, was taken to be uniform in the range $\left [ -1,1 \right ]$. The polarization angle $\psi$ was taken to be uniform in the range $\left [ 0,\pi \right ]$. The initial orbital phase, $\phi_0$, was taken to be uniform in the range $\left [ 0,2\pi \right ]$.
The gravitational wave emitted by a monochromatic source is calculated using the quadrupole approximation \cite{Landau:1962,Peters:1963}.
In this approximation, the gravitational wave signals are described as a combination of the two polarizations ($+, \times$)\citep{Korol:2021pun}: 
\begin{eqnarray}\label{eq:hplus_hcross}
    h_+(t) = \mathcal{A}(1+\cos\iota^2)\cos( \Phi(t)) \nonumber\\
    h_{\times}(t) = 2\mathcal{A}\cos\iota\sin(\Phi(t)).
\end{eqnarray}
In above expression,
\begin{eqnarray}
	\mathcal{A} =\frac{2(G\mathcal{M})^{5/3}}{c^{4}D_L}(\pi f)^{2/3}\nonumber\\
    \Phi(t)=2\pi ft + \pi\dot{f}t^2 + \phi_0  \nonumber\\
    \dot{f}=\frac{96}{5}  \pi^{8/3} \left( \frac{G{\cal M}}{c^3} \right)^{5/3} f^{11/3}. 
\end{eqnarray}
where $\mathcal{M}\equiv (m_1 m_2)^{3/5}/(m_1+m_2)^{1/5}$ is the chirp mass, and $D_L$ is luminosity distance, and $G$ and $c$ are the gravitational constant and the speed of light, respectively.
This is a source frame, but for detectors, Solar System Barycenter ($SSB$) frame, based
on the ecliptic plane, is selected. In $SSB$ frame, the standard spherical coordinates $\left(\theta,\phi\right)$ and the associated spherical orthonormal basis vectors $\left(\bf{e}r,\bf{e}\theta,\bf{e}\phi\right)$ are confirmed. The position of the
source in the sky is parametrized by the ecliptic latitude $\beta =\pi/2 - \theta$ and the ecliptic 
longitude $\lambda = \phi$. The gravitational wave propagation vector $\hat{k}$ in spherical coordinates is expressed as
\begin{equation}
    \hat{k} = -\bf{e} r = -\cos\beta\cos\lambda\,\hat{x} - \cos\beta\sin\lambda\,\hat{y} -
    \sin\beta\,\hat{z} \,.
\end{equation}
The reference polarization vectors is the following:
\begin{eqnarray}
    \hat{u} &=& -\bf{e}\theta = -\sin\beta\cos\lambda\,\hat{x} -\sin\beta\sin\lambda\,\hat{y} + \cos\beta\,\hat{z} \nonumber\\
    \hat{v} &=& -\bf{e}\phi = \sin\lambda\,\hat{x} - \cos\lambda\,\hat{y} .
\end{eqnarray}
The last degree of freedom between the frames corresponds to the rotation around the line of
sight, and is represented by the polarization angle $\psi$.
The polarization tensors are given by
\begin{eqnarray}
    {\mbox{\boldmath$\epsilon$}}^+ &=& \cos(2\psi) {\bf e}^+ - \sin(2\psi)
      {\bf e}^\times \nonumber\\
    {\mbox{\boldmath$\epsilon$}}^\times &=& \sin(2\psi) {\bf e}^+ +
    \cos(2\psi) {\bf e}^\times \, ,
\end{eqnarray}
where the
basis tensors ${\bf e}^+$ and ${\bf e}^\times$ are expressed in terms of two orthogonal unit vectors,
\begin{eqnarray}
    {\bf e}^+ &=& \hat{u} \otimes \hat{u} - \hat{v} \otimes \hat{v}
    \nonumber\\
    {\bf e}^\times &=& \hat{u} \otimes \hat{v} + \hat{v} \otimes \hat{u} \,.
\end{eqnarray}

An arbitrary gravitational wave traveling in the $\hat{k}$ direction can be written as the linear sum of two independent polarization states,
\begin{equation}\label{wave}
    {\bf h}(\xi) = h_+(\xi) {\mbox{\boldmath$\epsilon$}}^+
     + h_\times(\xi) {\mbox{\boldmath$\epsilon$}}^\times \,,
\end{equation}
where the wave variable $\xi = t - \hat{k} \cdot {\bf x}$ gives the surfaces of constant phase. 

\subsection{sky map of WD binary sources}\label{Sdwdb_sm}
Combining the theoretical model with the observations, the resulting new model may be more in line with the actual distribution. In the paper\cite{hol2018}, using Gaia DR2 data \cite{Gaia_2018}, an up-to-date sample of white dwarfs within 20 pc of the Sun is presented. So the WDs are distributed in the disc according to an exponential radial stellar profile with an isothermal vertical distribution \cite{Korol:2021pun}:  
\begin{equation} \label{eqn:dwd_positions} 
\rho(R,z)= \rho_{{\rm WD},\odot} \, e^{-\frac{R-R_\odot}{R_{\rm d}}} \sech^2 \left( \frac{z-z_\odot}{z_{\rm d}} \right),
\end{equation}
where $\rho_{{\rm WD},\odot} = (4.49 \pm 0.38) \times 10^{-3}$ pc$^{-3}$ is the local WD density estimated by \citep{hol2018} , $0 \le R \le 20$\, kpc is the cylindrical radial coordinate measured from the Galactic centre, $-2 \le z \le 2$\, kpc is the height above the Galactic plane, $R_{\rm d} = 2.5\,$kpc is the disc scale radius, and  $z_{\rm d}=0.3$\, kpc is the disc scale height \citep[e.g.][]{Juric2008,mac17}. Sun's position is set to $(R_\odot,z_\odot) = (8.1,0.03)$\,kpc \citep[e.g.][]{abu19}.

The total number of WD stars in the Milky Way disc by integrating the WD density profile is $(2.7 \pm 0.2) \times 10^{9}$.
To count the total number of WD binaries as gravitational wave sources, this number is multiplied by the WD fraction $f_{\rm WD} = 0.095 \pm 0.020$, derived by \cite{maoz18} for orbital separations $\lesssim 4$\, AU. The distribution of primary mass $m_1$ , is selected as a WD mass function that follows a three-component Gaussian mixture \cite{kep15} with means $\mu = \{0.65, 0.57, 0.81\}$\,\msun\,, standard deviations $\sigma = \{0.044, 0.097, 0.187\}$\,\msun\,, and respective weights $w = \{0.81, 014, 0.05\}$.

\begin{figure}
    \includegraphics[width=0.89\textwidth]{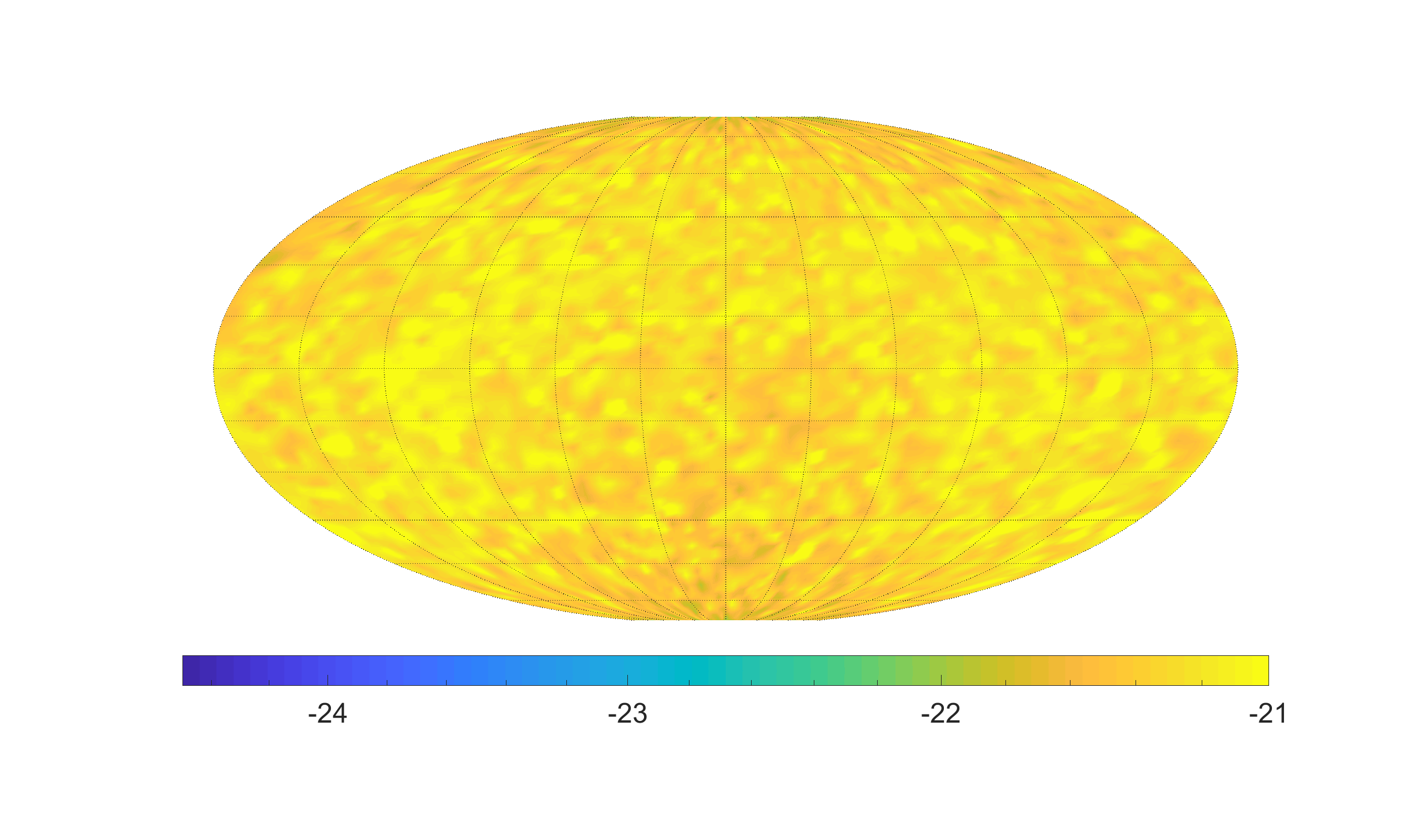}
    \caption{The sky map of amplitude intensity of the WD binary signal at different spatial positions. The horizontal axis represents the 
longitude ($\lambda \in \left[ - \pi, \pi\right]$)  and the vertical axis represents the ecliptice latitude ($\beta\in \left[ -\pi/2,\pi/2\right]$)}
     \label{fig:skymap1}
\end{figure}

Here, we will give a sky map of the amplitude of the WD binary signal sources. The coordinates of sky map  under the $SSB$-frame are composed of the spatial position coordinates ($\beta,\lambda$). The ecliptice latitude ($\beta\in \left[ -\pi/2,\pi/2\right]$) is divided into $m$ equal parts and the ecliptic 
longitude ($\lambda \in \left[ 0,2\pi\right]$) is divided into $n$ equal parts. 
So we can get the grids of  $m\times n$ on the sky map. Note that the horizontal axis represents the longitude  $\lambda \in \left[ - \pi, \pi\right]$. Then We calculate the amplitude of WD binaries based on the parameters according to the disc density distribution and the observationally motivated distributions of WD in section \ref{Ssm_dsm}. Because there are so many of WD binaries, there are many WD binary signals in each spatial grid. Here, we choose only the maximum value and keep it as the amplitude value at that spatial location. This sky map is shown in Fig.~\ref{fig:skymap1}. 
The sky map takes the sun as the observation center. For WD binaries in the whole space, the distribution of the amplitude intensity of the wave source in the sky map is not uniform. The location distribution of WD binaries in the Milky Way determines the intensity distribution of WD binary sources in the the whole sky.

\subsection{sky map of WD binaries projected on detectors}\label{Sds_sm}

\begin{figure*}
    \centering
    \vspace{-0.35cm}
    \subfigtopskip=2pt
    \subfigbottomskip=2pt
    \subfigcapskip=-5pt
    \subfigure[$\kappa=0^{\circ}$]{
    \includegraphics[width=0.45\textwidth]{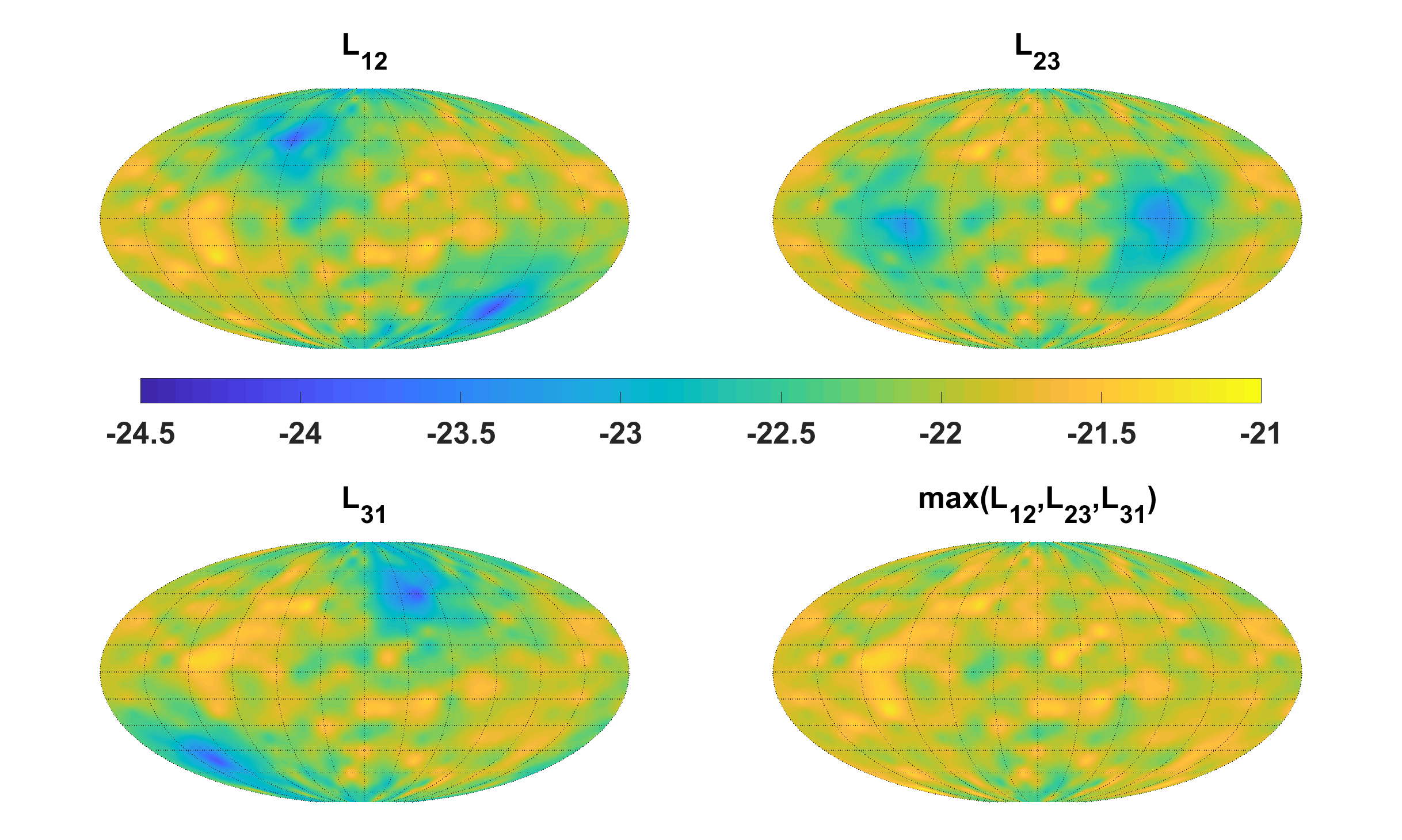}
    }
    \quad
    \subfigure[$\kappa=45^{\circ}$]{
    \includegraphics[width=0.45\textwidth]{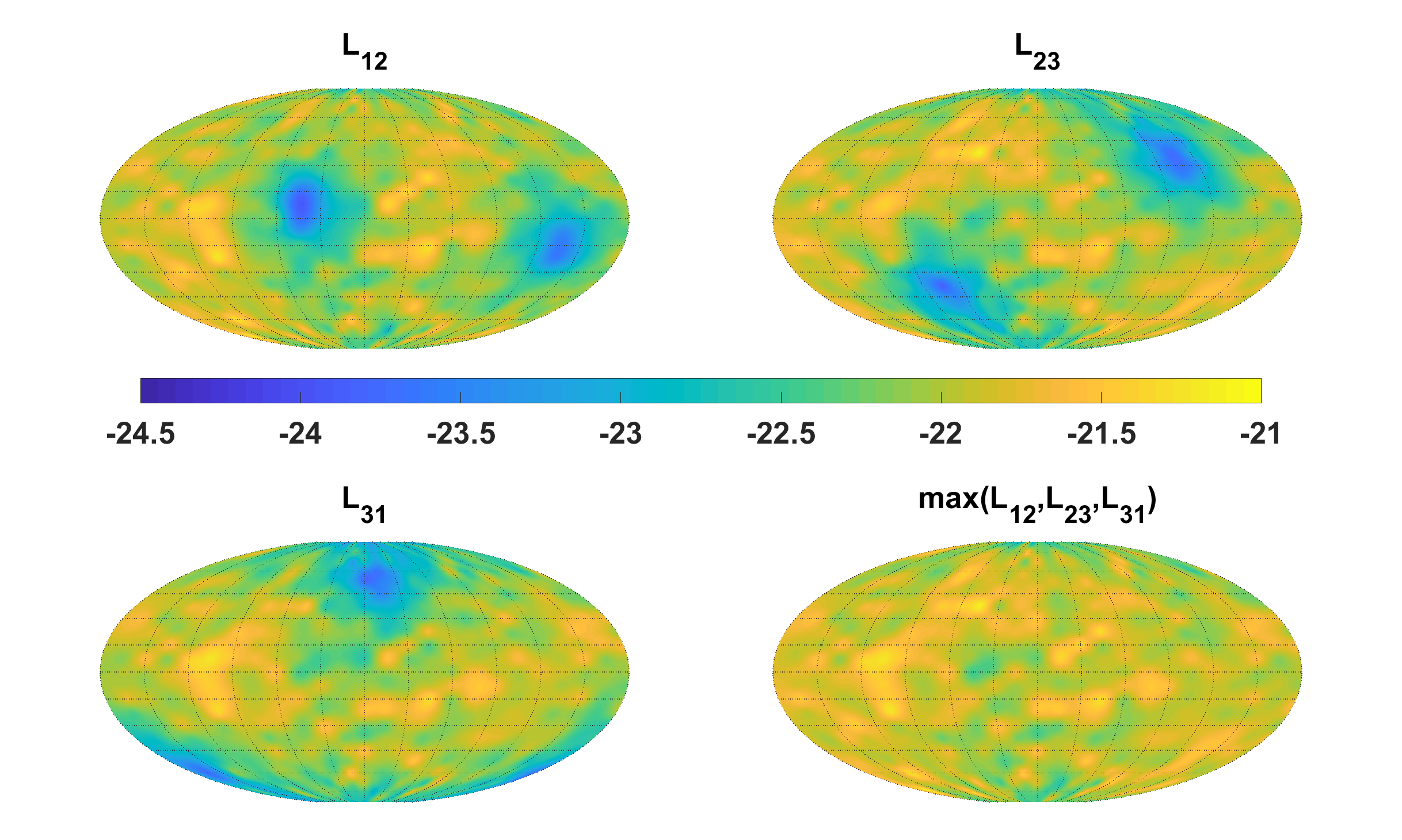}
    }
   
    \subfigure[$\kappa=90^{\circ}$]{
    \includegraphics[width=0.45\textwidth]{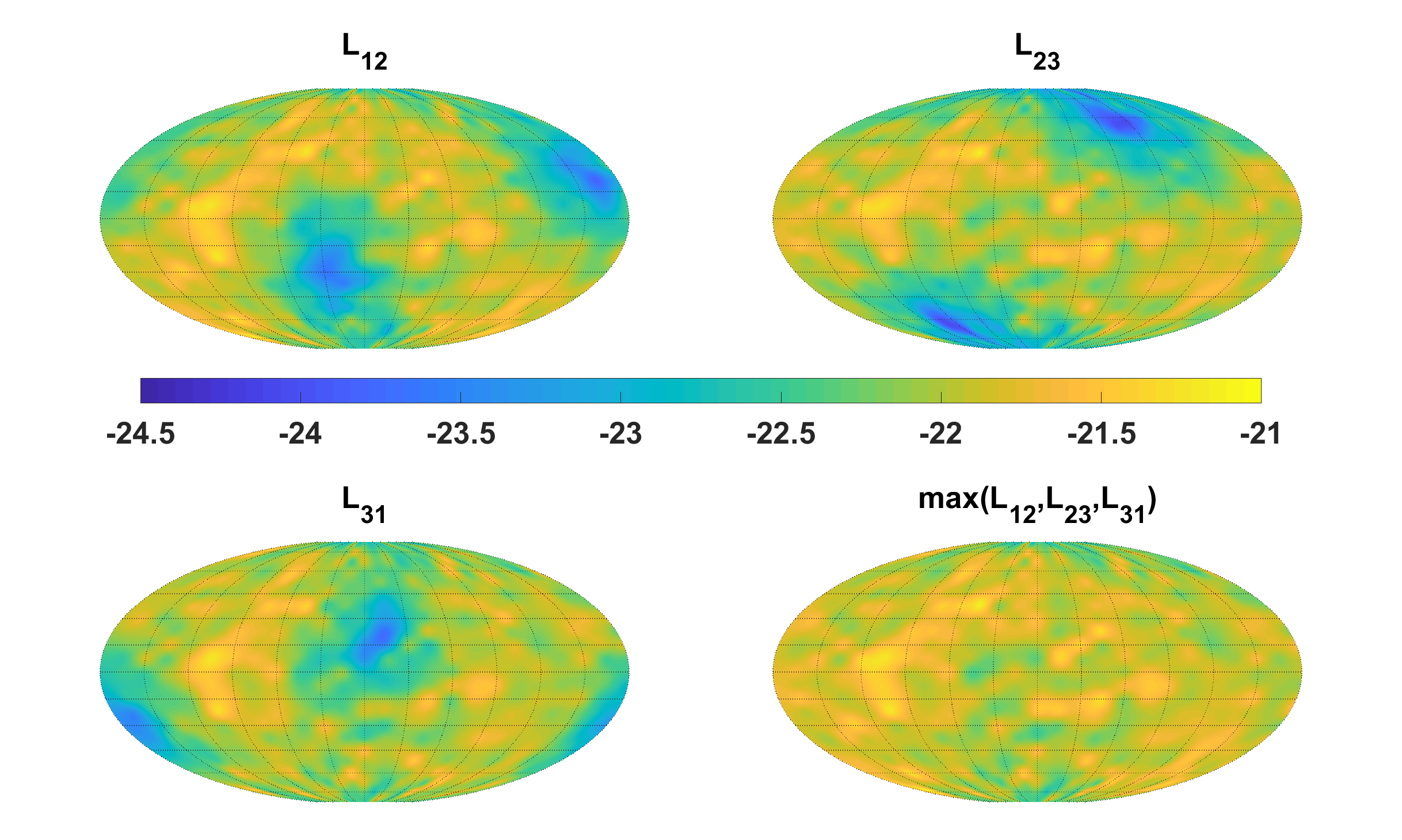}
    }
     \quad
    \subfigure[$\kappa=135^{\circ}$]{
    \includegraphics[width=0.45\textwidth]{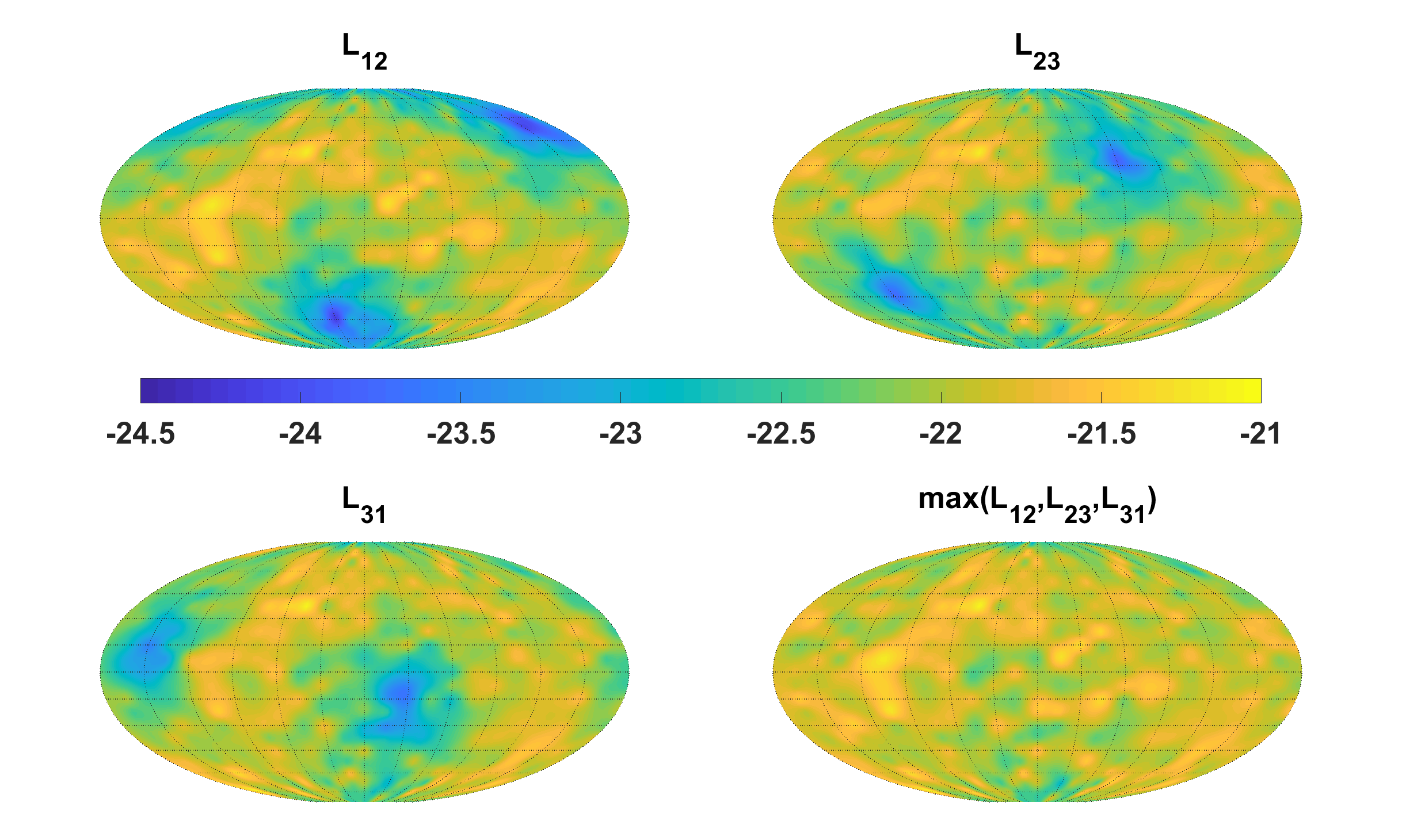}
    }
     \caption{The sky map of amplitude of the source projected on $L_{12}$, $L_{23}$, and $L_{31}$ and the maximum amplitude of the WD binaries projected on $L_{12}$, $L_{23}$, and $L_{31}$ for different variable $\kappa$ respectively. The variable $\kappa$ ($\kappa \in \left[ 0,2\pi\right]$) is divided into 8 equal parts.}
     \label{fig:skymap_hij_1}
\end{figure*}    

\begin{figure*}
    \centering
    \vspace{-0.35cm}
    \subfigtopskip=2pt
    \subfigbottomskip=2pt
    \subfigcapskip=-5pt
    \subfigure[$\kappa=180^{\circ}$]{
    \includegraphics[width=0.45\textwidth]{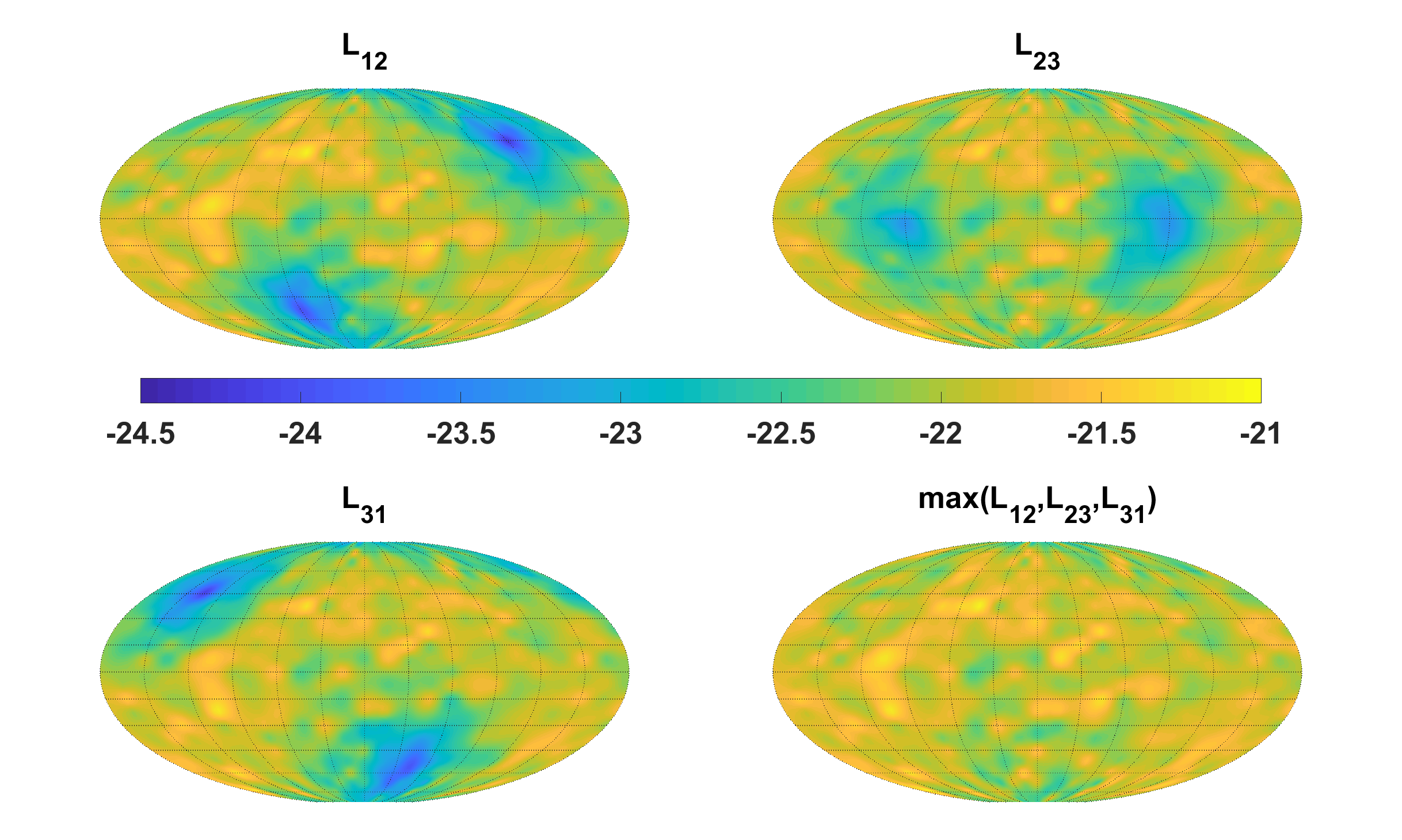}
    }
    \quad
    \subfigure[$\kappa=225^{\circ}$]{
    \includegraphics[width=0.45\textwidth]{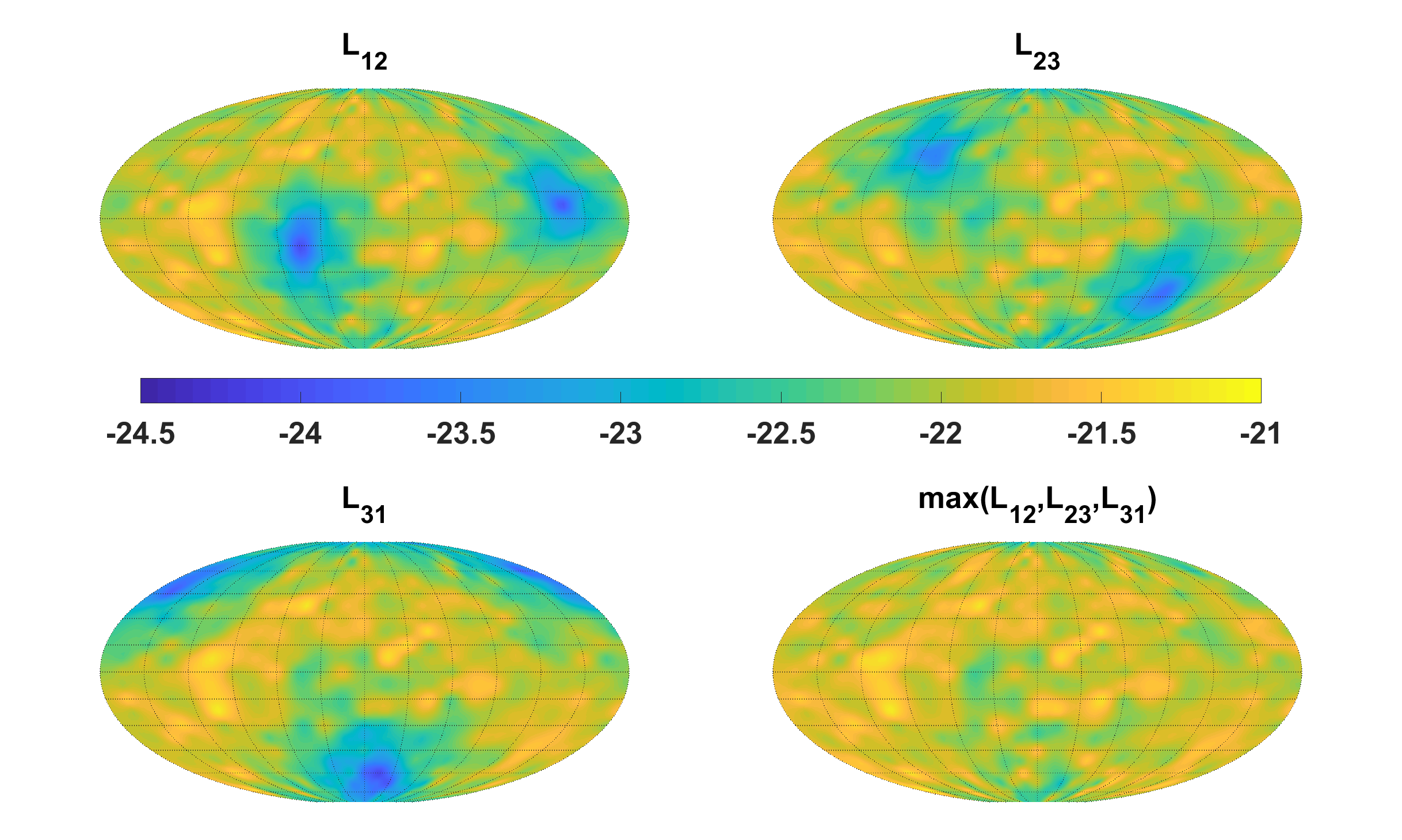}
    }
   
    \subfigure[$\kappa=270^{\circ}$]{
    \includegraphics[width=0.45\textwidth]{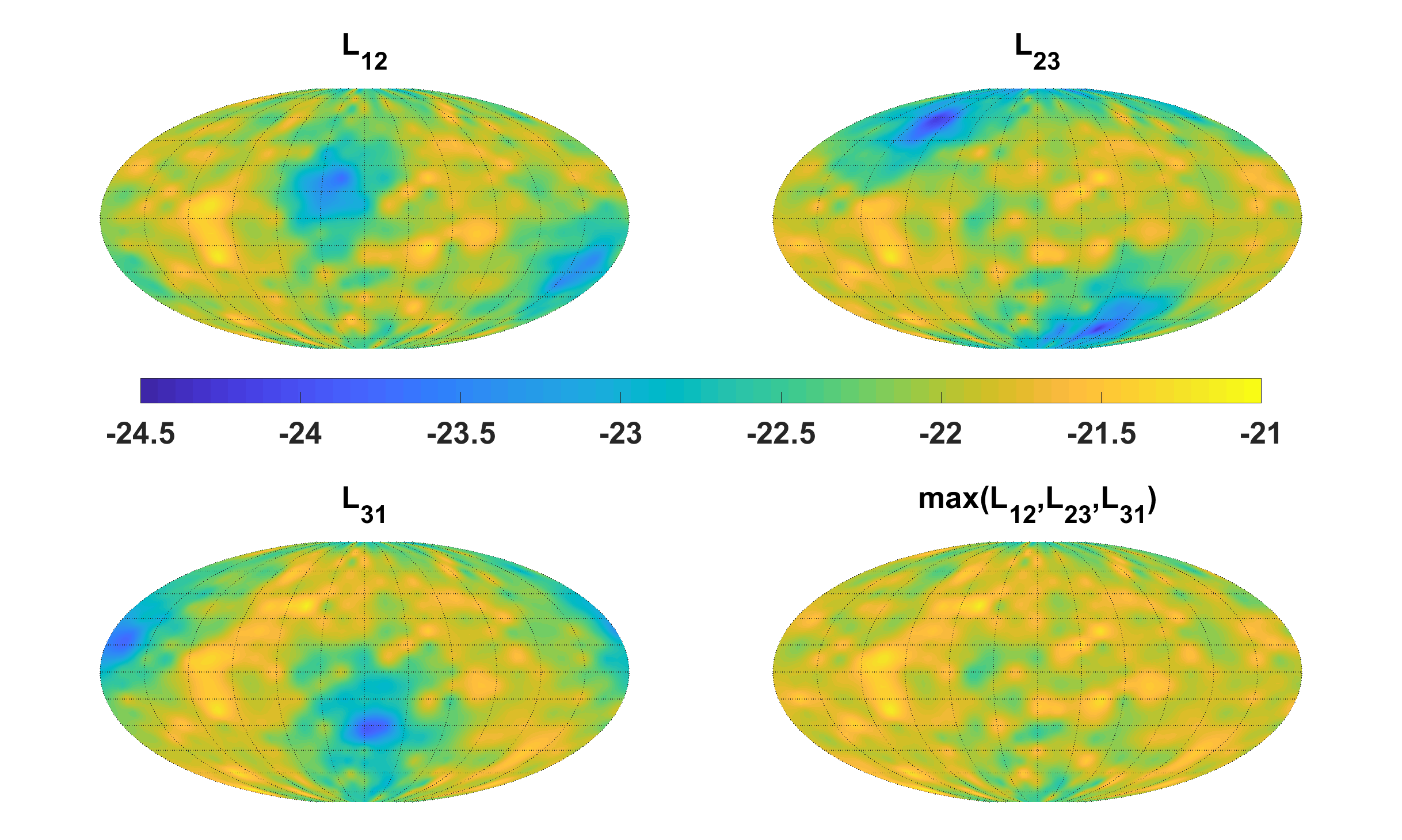}
    }
     \quad
    \subfigure[$\kappa=315^{\circ}$]{
    \includegraphics[width=0.45\textwidth]{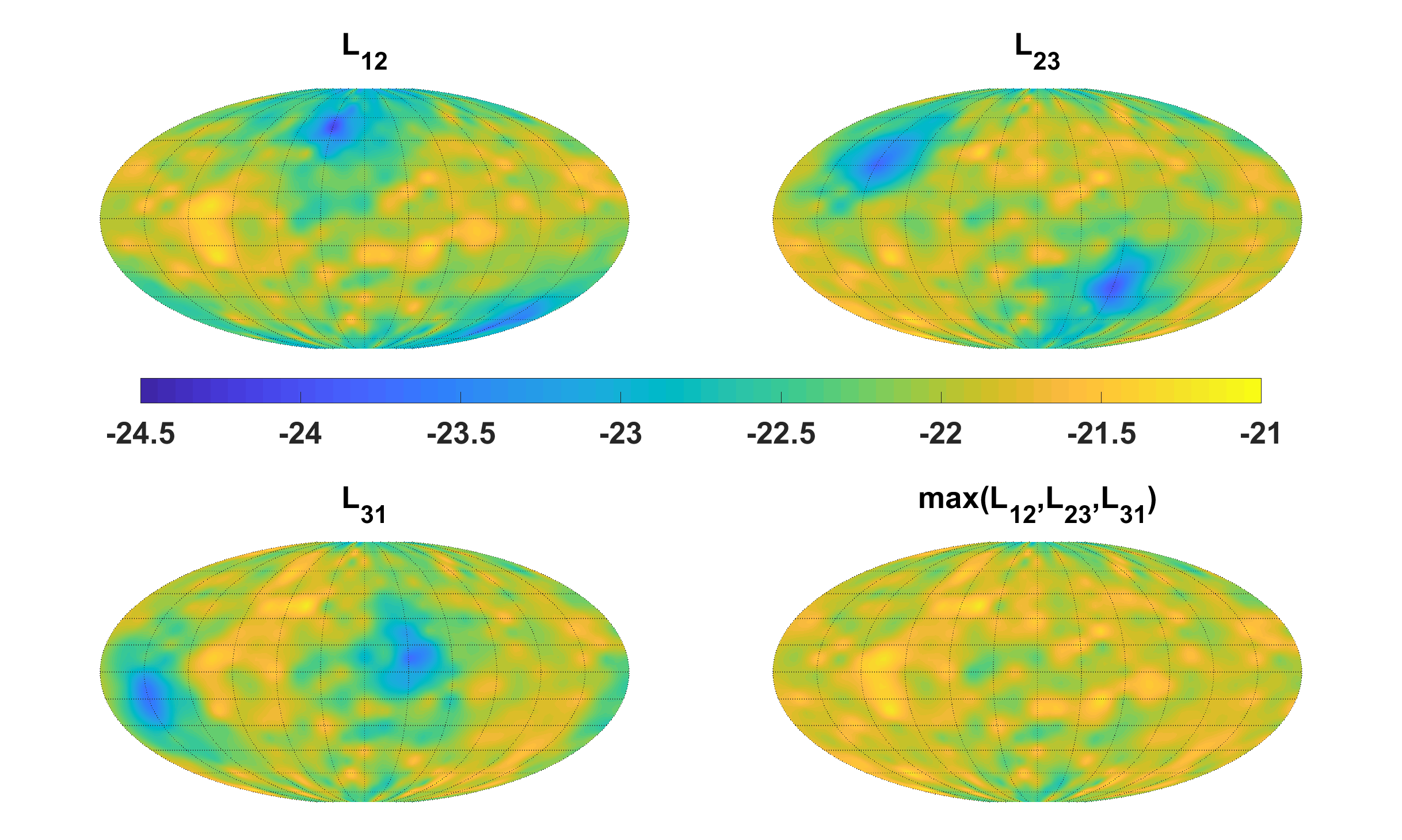}
    }
     \caption{The sky map of amplitude of the source projected on $L_{12}, L_{23}, L_{31}$ and the maximum amplitude of the WD binaries projected on $L_{12}, L_{23}$ and $L_{31}$ for different variable $\kappa$ respectively. The variable $\kappa$ ($\kappa \in \left[ 0,2\pi\right]$) is divided into 8 equal parts.}
     \label{fig:skymap_hij_2}
\end{figure*}    
For an identified wave source, in order to better observe the effect of different orbital positions of the detectors on its observation data, we just consider the short-time observation of a WD binary. Since its observation time is so small relative to the one-year observation period, the change of amplitude of the projected signal on detectors caused by the the detector motion in the short  observation time can be ignored. Here, the amplitude of the projected signal is normalized and represents the maximum value of the projected signal in a period. 
For the same source and the same short observation time, the different detector positions in a one-year period can get the different amplitude of the projected signal on detectors. In contrast to the section \ref{Sds_ga}, the variable $\kappa$ ($\kappa \in \left[ 0,2\pi\right]$) is divided into 8 equal parts and we can get 8 sky maps for amplitude of WD binaries projected on each detection arm. The sky map shows amplitude of WD binaries projected on $L_{12}, L_{23}, L_{31}$ in the whole space projected on three detection arms respectively. 
The coordinates under the $SSB$-frame are also composed of the spatial position coordinates ($\beta,\lambda$). In every sky map, the horizontal axis represents the 
longitude ($\lambda \in \left[ - \pi, \pi\right]$)  and the vertical axis represents the ecliptice latitude ($\beta\in \left[ -\pi/2,\pi/2\right]$). So we can get the grids of  $24\times 24$ on the sky map. Then we calculate the amplitude maximum of WD projected signal for each grid on sky map. Here, the parameters of the WD binaries are the same as the one in section \ref{Sdwdb_sm} and the detector response is shown in Eq.~(\ref{deltaL2}). Here, we choose only the maximum value and keep it as the amplitude maximum value at that spatial location. The sky map of amplitude of the source projected on $L_{12}$, $L_{23}$, and $L_{31}$ and the maximum amplitude of the WD binaries projected on $L_{12},L_{23}$ and $L_{31}$ for different variable $\kappa$ respectively are shown in Fig.~\ref{fig:skymap_hij_1} and  Fig.~\ref{fig:skymap_hij_2}. For a WD binary of the frequency  $f_0$, the sampling frequency $F_s = 12.8 \times f_0$, the data counts $n = 128$. So we can calculate the observation time $T_{OB}=n/F_s =10/ f_0$. For the frequency $f_0 = 10^{-4} Hz$, the observation time $T_{OB} = 781.25 s$ can be calculated, which is a short-time observation for the one-year period of detector orbit. Due to the different frequencies of WD binary signals, the observation time of each WD binary signal is not the same. However, the amplitude value does not change due to the difference in the number of observation signal periods caused by the slight difference in the short observation time.

\begin{figure*}
    \centering
    \vspace{-0.35cm}
    \subfigtopskip=2pt
    \subfigbottomskip=2pt
    \subfigcapskip=-5pt
    \subfigure[VB: J0806]{
    \includegraphics[width=0.47\textwidth]{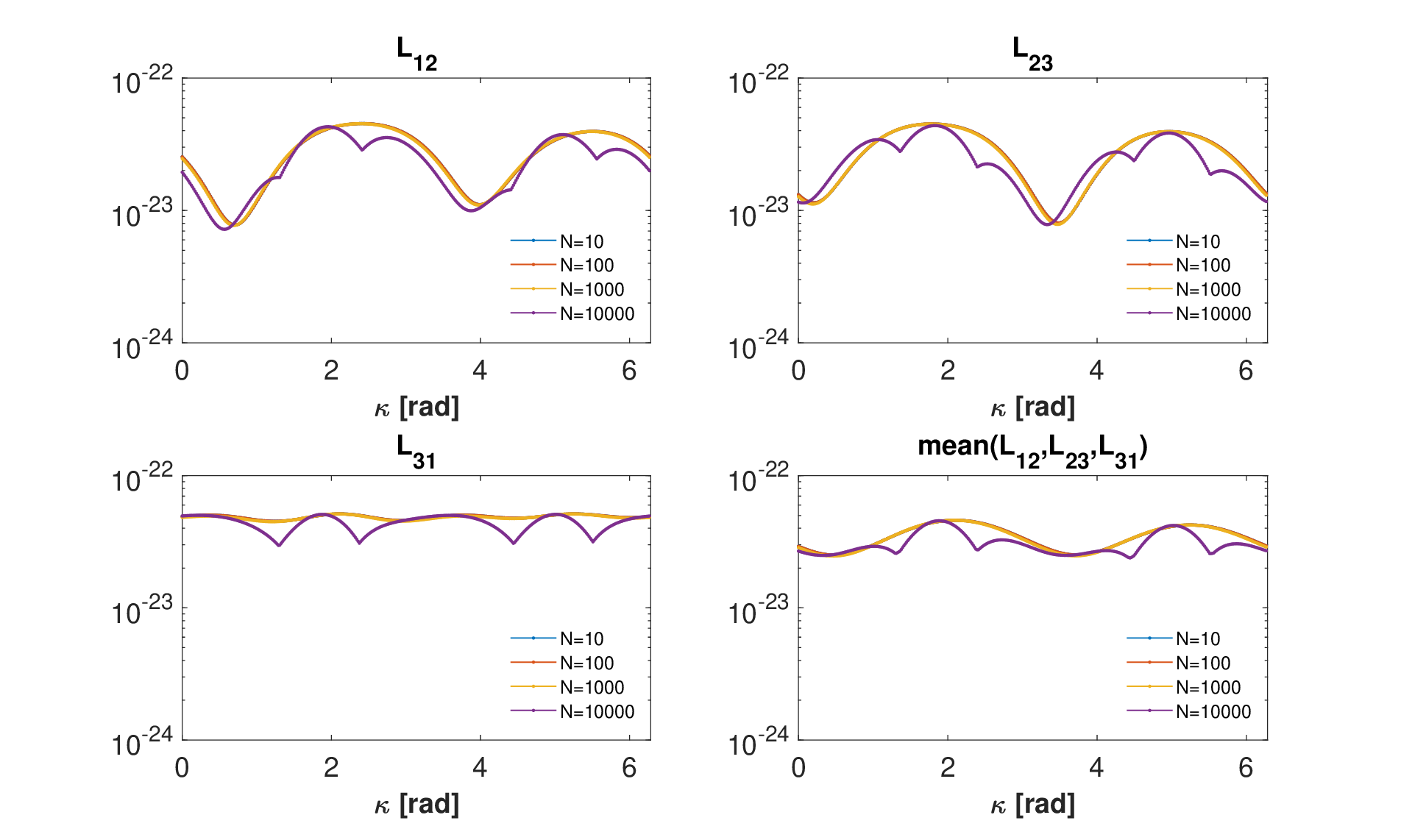}
    }
    \quad 
    \subfigure[VB: V407 Vul]{
    \includegraphics[width=0.47\textwidth]{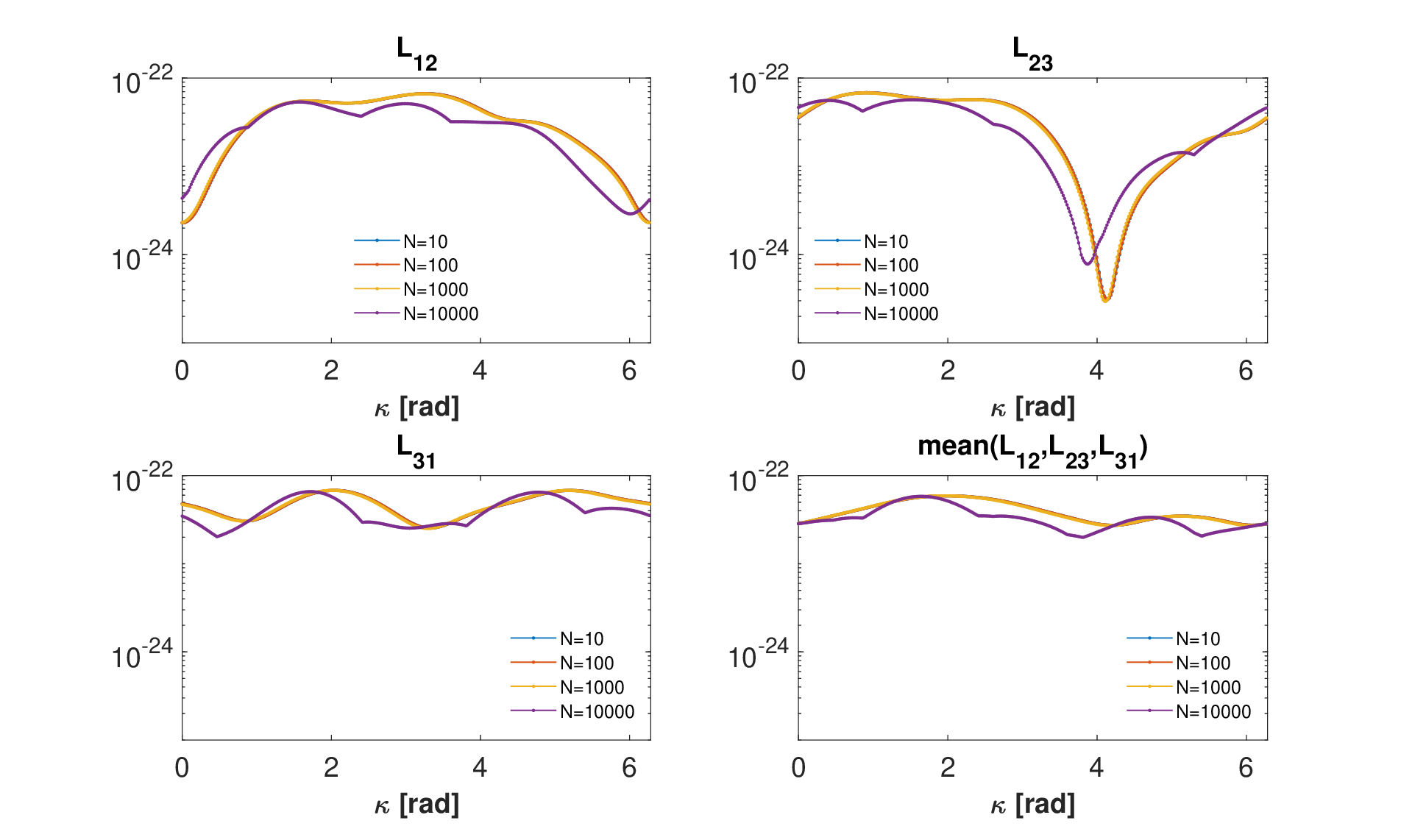}
    }
   
    \subfigure[VB: ES Cet]{
    \includegraphics[width=0.47\textwidth]{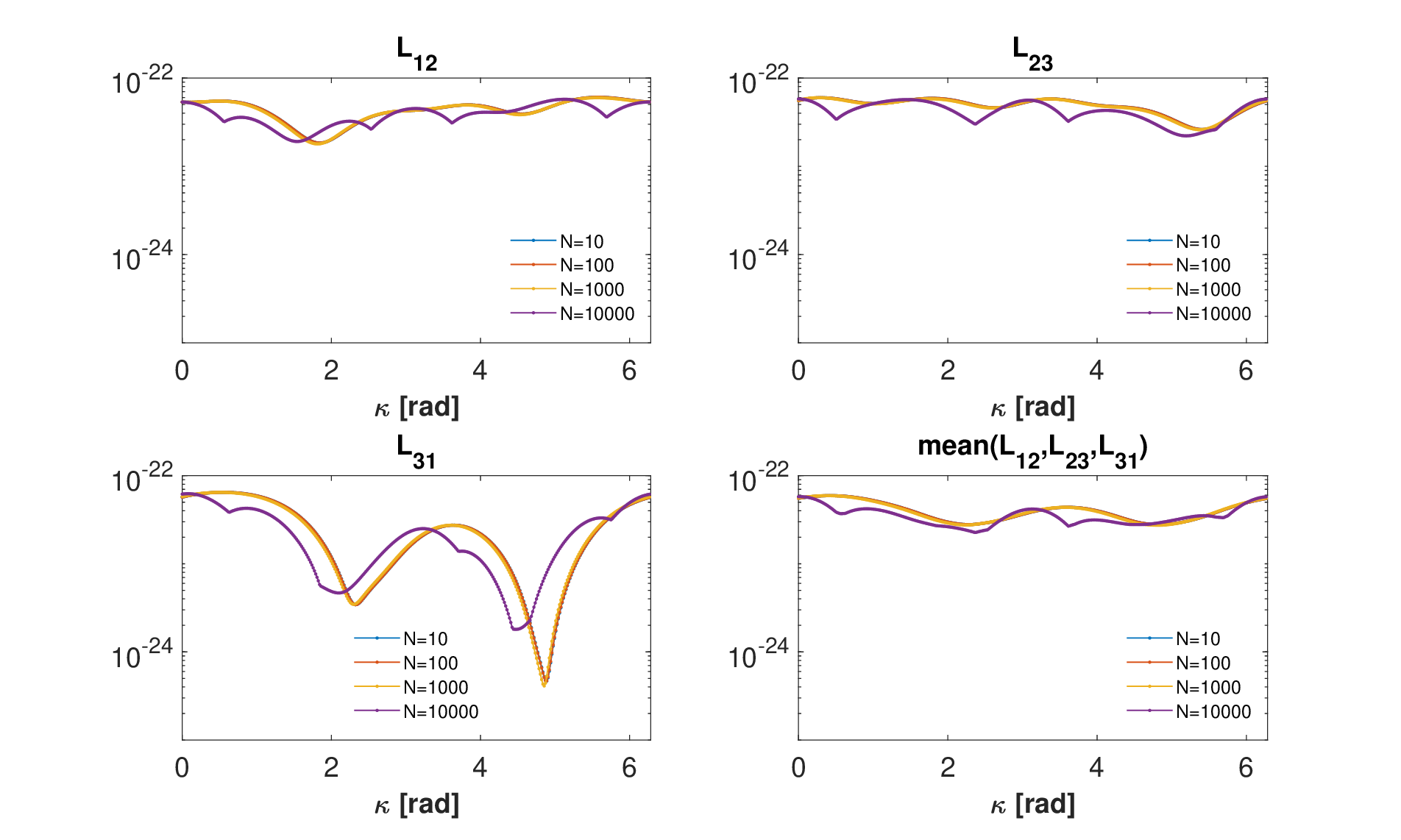}
    }
     \quad
    \subfigure[VB: SDSSJ1351]{
    \includegraphics[width=0.47\textwidth]{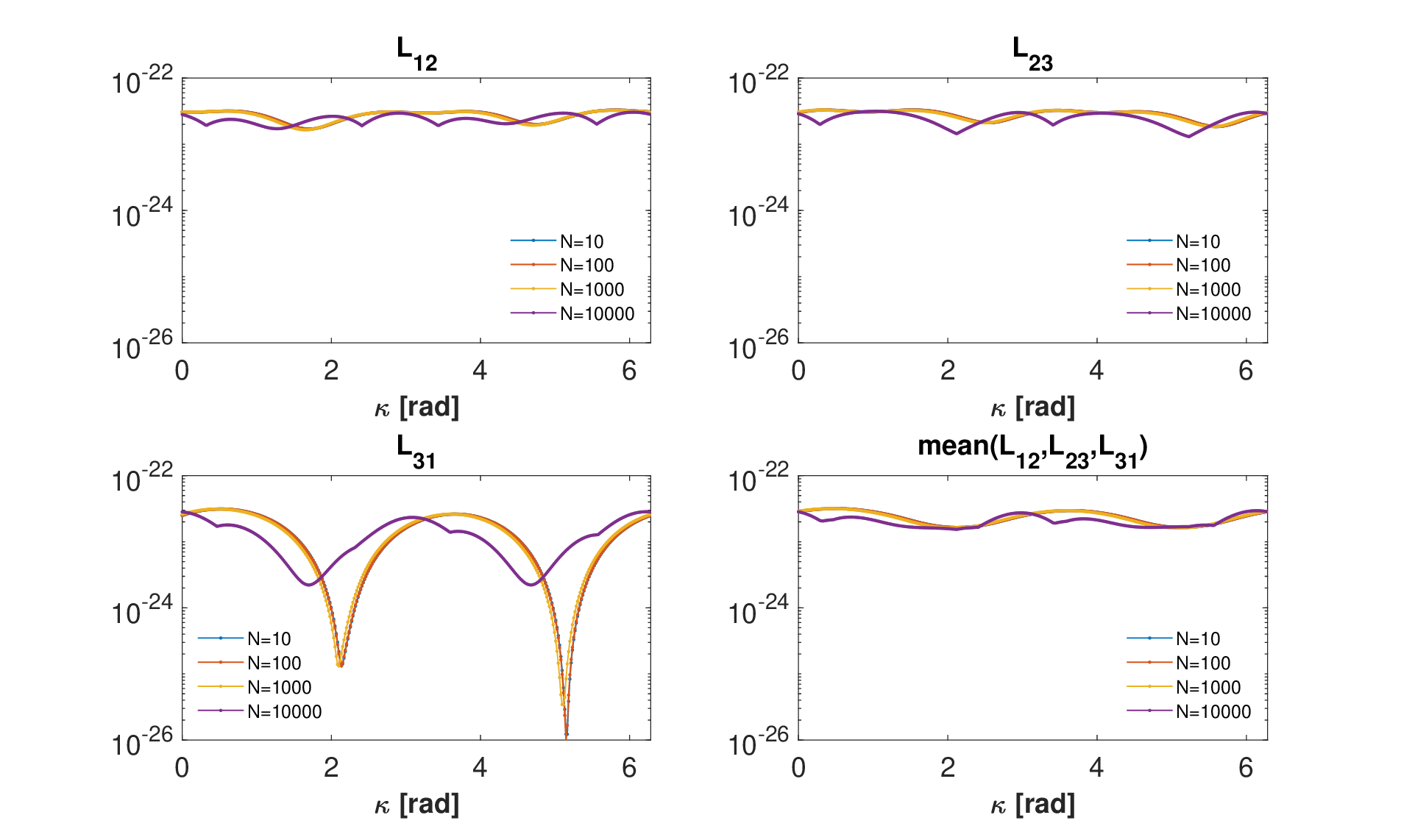}
    }
     \caption{The amplitude of projection signals of four VB signals at different observation time. N is the times of GW period observed during the observation time.}
     \label{fig: VB_A_Tob}
\end{figure*}

Next, we discuss the effect of observation time on the amplitude of the projected signals. We select 4 verification binaries signals: J0806, V407 Vul, ES Cet, SDSSJ1351, whose parameters are detailed in \cite{Kupfer:2018jee}. The polarisation angle is set $\psi=0$, and initial orbital phase is set $\phi_0=0$. Then we can calculate the amplitudes of the projected signals on the three detection arms $L_{12}$, $L_{23}$, and $L_{31}$, respectively, and obtain the results of the joint observation of the three detection arms. 
We selected 4 different observation time $T_{OB} = N/\times f_0$. Where $f_0$ is the frequency of VB signal and N is the number of the VB cycles observed during the observation time. We set $N=10, 100, 1000, 10000$. The amplitude of projection signals of four VB signals at different observation time are shown in Fig.~\ref{fig: VB_A_Tob}.
From the Fig.~\ref{fig: VB_A_Tob}, we can find that the amplitude difference is not significant at $N=10, 100, 1000$. However, the amplitude is significantly different at $N=10000$.

\begin{figure}
    \centering
    \includegraphics[width=0.89\textwidth]{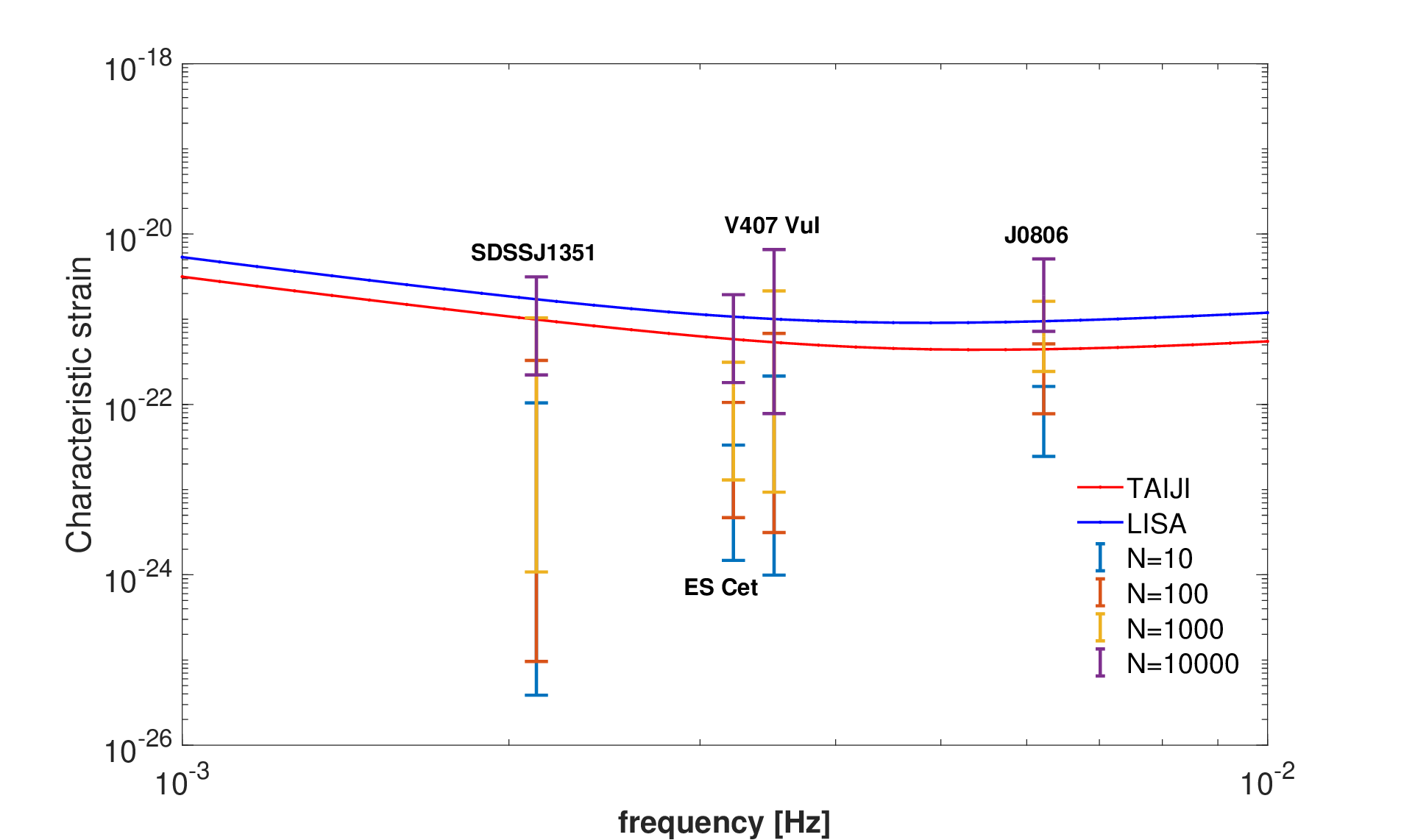}
    \caption{The variation range of the characteristic strain $h_c$ of the 4 VBs for all $L_{{ij},\kappa}$ for TAIJI (The 4 color error bars correspond to 4 different observation times and N is the number of the VB cycles observed during the observation time) compared with the noise amplitude $h_n$ of TAIJI (red line) and LISA (blue line).}
    \label{fig:VB_hc_bar_Tob}
\end{figure}

In the Fig.~\ref{fig:VB_hc_bar_Tob}, we give the variation range of the characteristic strain $h_c$ of the 4 VB signals for all $L_{{ij},\kappa}$ for TAIJI (The 4 color error bars correspond to 4 different observation times and N is the number of the VB cycles observed during the observation time) compared with the noise amplitude $h_n$ of TAIJI (red line) and LISA (blue line). Here, the characteristic strain $h_c=\sqrt{N}\times A$, the noise amplitude $h_n = \sqrt{f\times S_n(f)}$. 

\section{Localizing the sky position of identified WD binary}\label{sec:lsp}

Due to the periodic orbit motion of the detector, the results of short time observation are related to the orbit position of the detector. For some specific WD binary sources, the best localizing the sky position of the sources at the best detector position is achieved.

Here, we select 4 verification binaries signals: J0806, V407 Vul, ES Cet, SDSSJ1351 as the identified sources, whose parameters are detailed in \cite{Kupfer:2018jee}. The polarisation angle is set $\psi=0$, and initial orbital phase is set $\phi_0=0$.
The sky location parameters (\mbox{$\lambda [rad], sin \beta, D_L [kpc]$}) of 4 VB signals are set as: (2.1021, -0.0820, 5), (5.1486, 0.7288, 1.786), (0.4295, -0.3475, 1.584), (3.6370, 0.0780, 1.317), respectively.
The posterior distribution lies typically in a small volume of the parameter space. The research on localizing the sky position of identified WD binary based on short-term period observation is mainly divided into 3 steps:

$\textbf{Step 1}$: Find the reduced parameter space as the new prior. Estimate $1 \sigma$ of the sky position parameters using the Fisher Information Matrix and frame the $3 \sigma$ covariance region centered on the true value of the parameter as the reduced parameter space for the next steps. 

$\textbf{Step 2}$: Find best orbit position of detector. For each VB source, find best orbit position of detector with the smallest parameter error. 

$\textbf{Step 3}$: Get the posterior. Use Metropolis-Hastings MCMC to get the posterior distribution of the location parameters at the best orbit position of detector.

In this paper, the detector response form in the time domain is used, so we will get the gravitational wave data in the time domain. When calculating the likelihood for each samples, we convert the data in time domain into frequency-domain through FFT.
 For a WD binary of the frequency  $f_0$, the sampling frequency $F_s = 10 \times f_0$, the data counts $n = 128$. So we can calculate the observation time $T_{OB}=n/F_s =1/(12.8\times f_0)$. 
 Detailed calculation and related theories are shown below.
\subsection{Bayesian inference}\label{sec:Bayes}
The Bayesian inference is based on 
calculating the  posterior probability distribution function (PDF) of  
the unknown parameter set 
$\boldsymbol{\theta}=\{\theta_{1},\dots,\theta_{m}\}$ in a given model,
which actually updates our state of belief from the prior PDF of $\boldsymbol{\theta}$ 
after taking into account the information provided by  the experimental data set $D$.
The posterior PDF is related to the prior PDF by  the Bayes's therom
\begin{align}\label{eq:Bayes}
p(\boldsymbol{\theta}|D)=\frac{\mathcal{L}(D|\boldsymbol\theta)\pi(\boldsymbol\theta)}{p(D)} ,
\end{align}
where $\mathcal{L}(D|\boldsymbol\theta)$ is the likelihood function,
and 
$\pi(\boldsymbol\theta)$ is the prior PDF which 
encompasses our state of knowledge on the values of the parameters before 
the observation of  the data.
The quantity $p(D)$ is the Bayesian evidence 
which is obtained by integrating the product  of the likelihood and the prior over
the whole volume of the parameter space
\begin{equation}
    p(D)=\int_{V} \mathcal{L}(D|\theta)\pi(\boldsymbol\theta) d\theta .
\end{equation}

The evidence is an important  quantity for Bayesian model comparison.It is straight forward to obtain the marginal PDFs of interested parameters 
$\{\theta_{1}, \dots, \theta_{n}\} (n<m) $ by integrating out other nuisance parameters $\{\theta_{n+1}, \dots, \theta_{m}\}$
\begin{equation}
    p(\theta_{1},\dots,\theta_{n})_{\text{marg}}=\int p(\boldsymbol\theta|D) \prod_{i=n+1}^{m}d\theta_{i} .
\end{equation}
The marginal PDF  is often used in visual presentation. 	
If there is no preferred value of $\theta_{i}$ in the allowed range ($\theta_{i,\text{min}}$, $\theta_{i,\text{max}}$),
the priors can be taken as a flat distribution
\begin{align}\label{eq:priors}
\pi(\theta_{i}) \propto
\left\{
\begin{tabular}{ll}
1, &  \text{for } $\theta_{i,\text{min}}<\theta_{i}<\theta_{i,\text{max}}$
\\
0, & \text{otherwise}
\end{tabular}
\right. 
.
\end{align}
The likelihood function is often assumed to be Gaussian
\begin{align}
\mathcal{L}(D|\boldsymbol\theta)=
\prod_{i}
\frac{1}{\sqrt{2\pi \sigma_{i}^{2}}}
\exp\left[
	-\frac{(f_{\text{th},i}(\boldsymbol\theta)-f_{\text{exp},i})^{2}}{2\sigma_{i}^{2}}  
\right]  ,  
\end{align}
where $f_{\text{th},i}(\boldsymbol\theta)$ are 
the predicted $i$-th observable from the model which 
depends on the parameter set $\boldsymbol\theta$, 
and
$f_{\text{exp},i}$ are the ones measured by the experiment with uncertainty $\sigma_{i}$.
For experiments with only a few events observed, 
the form of the likelihood function can be taken as Poisson. 
When the form of the likelihood function is specified, 
the posterior PDF can be determined by sampling the distribution 
according to the prior PDF and the likelihood function	
using  Markov Chain Monte Carlo (MCMC) methods.
The statistic mean value of a parameter
$\theta$ can be obtained from the posterior PDF 
$P(\boldsymbol\theta|D)$ 
in a straight forward manner.
Using the MCMC sequence 
$\{\theta^{(1)}_{i}, \theta^{(2)}_{i},\dots,
\theta^{(N)}_{i}\}$ 
of the parameter $\theta_{i}$
with $N$ the length of the Markov chain,
the mean (expectation) value $\langle \theta_{i} \rangle$ is given by

\begin{eqnarray}
    \langle \theta_{i} \rangle
    =
    \int \theta_{i} P(\theta_{i}|D) d\theta_{i}
    =\frac{1}{N}\sum_{k=1}^{N} \theta^{(k)}_{i}.
\end{eqnarray}
    
The $1\sigma$ standard  deviation of the parameter $\theta_{i}$ is given by
$\sigma^{2}=\sum^{N}_{k=1} (\theta^{(k)}_{i}-\langle \theta_{i} \rangle)^{2}/(N-1)$.

\subsection{Fisher Information Matrix}
The Fisher Information Matrix (FIM) is a popular choice to estimate the parameter uncertainty for GW signals.
Here, we choose joint observations of the detector $arm12$, $arm23$ and $arm31$ to calculate the FIM
\begin{equation}
    F_{ij} =  F_{ij}^{12}+F_{ij}^{23}+F_{ij}^{31}
\end{equation}
\begin{equation}
    F_{ij}^{lk} = \left \langle \frac{\partial h_{lk}}{\partial \theta^i}\bigg|\frac{\partial h_{lk}}{\partial \theta^j} \right \rangle 
\end{equation}
where $h_{lk}$ is the response of the detector $arm_{lk}$ to a gravitational wave signal and $\theta^i$ is $i_{th}$ component of the sky position parameters of the signal, $\{ \lambda, sin\beta, D_L\}$.

The instrument noise of detector was ignored, and we focus on the projected signal on the single arm detector of an identified WD binary. In this paper, $h_{lk}$ are the ones measured by the experiment with uncertainty $\sigma= a\times h_{lk}$.
So the noise is set as the Gaussian noise with mean $\mu =0$, standard deviation $\sigma= a\times h_{lk}$. 

The matrix inversion of the FIM gives the Gaussian covariance matrix, $F_{ij}^{-1}=C_{ij}$. 
To estimate the position of the WD binary in the sky, following \cite{Cutler:1997ta}, the solid angle 
\begin{equation}
    \Delta \Omega_s = 2\pi\cos{\beta}\sqrt{C_{\beta\beta}C_{\lambda\lambda}-C_{\beta\lambda}^2}
\end{equation}
where 
\begin{eqnarray}
    C_{\beta\beta}&=&C_{sin\beta sin\beta}\times \cos{\beta_0}^2\nonumber\\
    C_{\beta\lambda}&=&C_{sin\beta \lambda}\times \cos{\beta_0}
\end{eqnarray}
where $\beta_0$ is the real value of the parameter $\beta$ of the identified WD binary.

\subsection{best and worst orbit position for observation}
In this work, a research on localizing the sky position of identified DWD source based on short-term period observation is realized. Due to the periodic orbit motion of the detector, the results of short-term observation are related to the orbit position of the detector. For some identified DWD sources, the best localizing the sky position of the sources at the best detector position is achieved in this work.

\begin{figure}
    \centering
    \includegraphics[scale=0.55]{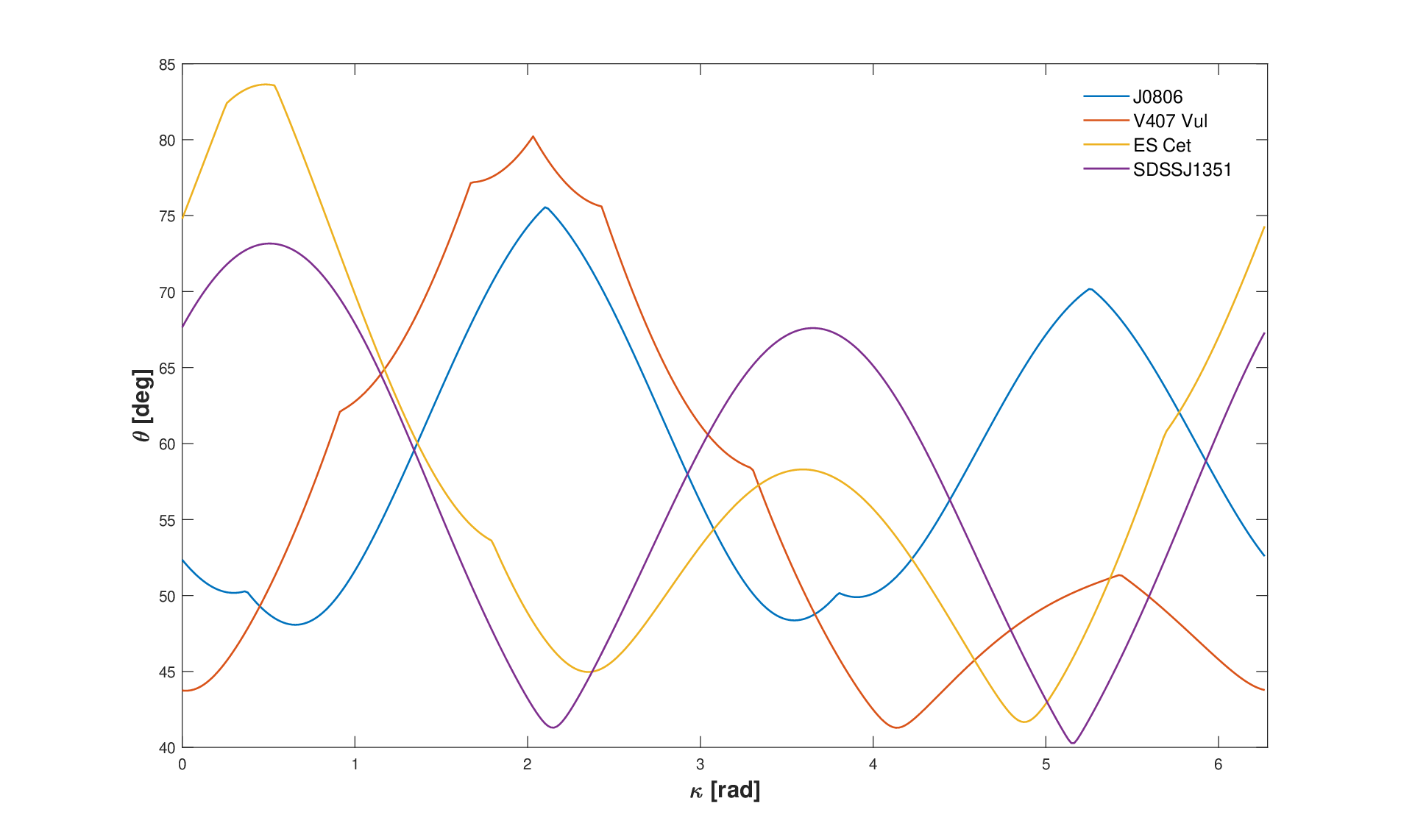}
    \caption{The relationship between the orbital position $\kappa$ and the mean of the Angle $\theta$ of $L_{12,\kappa}$, $L_{23,\kappa}$ and $L_{31,\kappa}$ for J0806, V407 Vul, ES Cet, SDSSJ1351,respectively. }
    \label{fig:mean3L_theta_kappa_4VB}
\end{figure}

Here, we select 4 verification binaries signals: J0806, V407 Vul, ES Cet, SDSSJ1351 as the identified sources. The sky location parameters ($\lambda [rad], sin \beta, D_L [kpc]$) of 4 VB signals are set as: (2.1021, -0.0820, 5), (5.1486, 0.7288, 1.786), (0.4295, -0.3475, 1.584), (3.6370, 0.0780, 1.317), respectively.
In Sec.~\ref{Sds_ga}, the annual modulation effect is mainly caused by the change of the Angle between the propagation direction of the GW source and the detector arm during the one-year orbit. The change of the Angle $\theta$ reflects the annual modulation effect. So we can calculate the relationship between the orbital position $\kappa$ and the mean of the Angle $\theta$ of $L_{12,\kappa}$, $L_{23,\kappa}$ and $L_{31,\kappa}$ for any DWD source that determines the sky location.
Here, take 365 different $\kappa$ values uniformly in the interval $\kappa \in \left[ 0,2\pi\right]$. In the Fig.~\ref{fig:mean3L_theta_kappa_4VB}, we give the relationship between the orbital position $\kappa$ and the mean of the Angle $\theta$ of $L_{12,\kappa}$, $L_{23,\kappa}$ and $L_{31,\kappa}$ for J0806, V407 Vul, ES Cet, SDSSJ1351, respectively. So we can get the best and worst observation location $\kappa$ for 4 VB sources. 
So localizing the sky position of 4 VB sources at the best and worst observation location $\kappa$ can be achieved.

\subsection{Posterior}\label{sec:Bse_slpe}
 
Sky location parameter is set by $\boldsymbol\theta=\{ \lambda, sin \beta, D_L\}$. 
The reduced parameter space as new prior was obtained at best and worst orbit position.
The reduced parameter space can be used to set the desired parameter boundaries as the new prior.
$136000$ samples of one MCMC chain were selected and then Metropolis-Hastings sampling was used to obtain the posterior distribution of parameters.
 
\begin{figure*}
\begin{minipage}[t]{0.49\linewidth}
    \centering
    \includegraphics[scale=0.58]{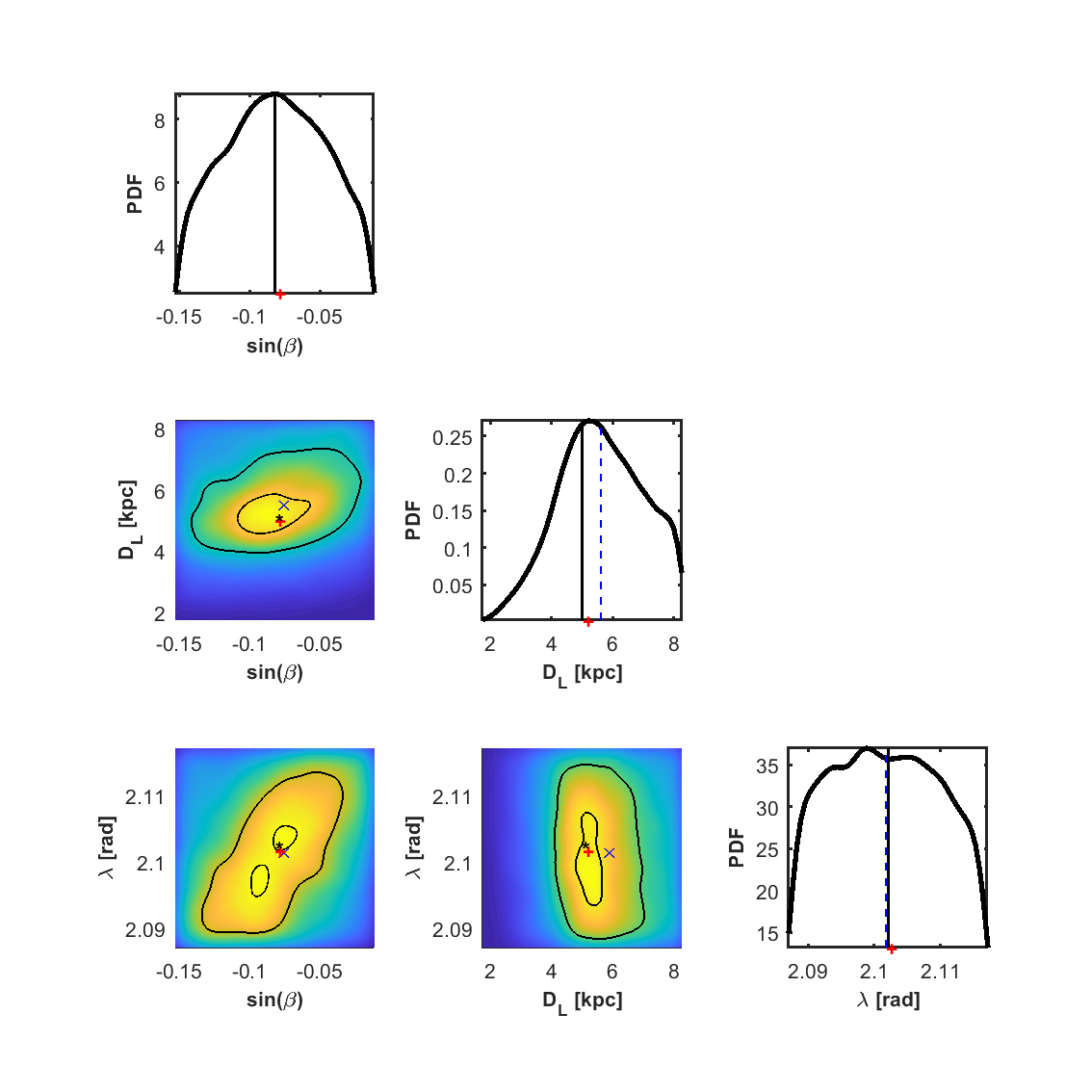}
\end{minipage}
\begin{minipage}[t]{0.49\linewidth}
    \centering
    \includegraphics[scale=0.58]{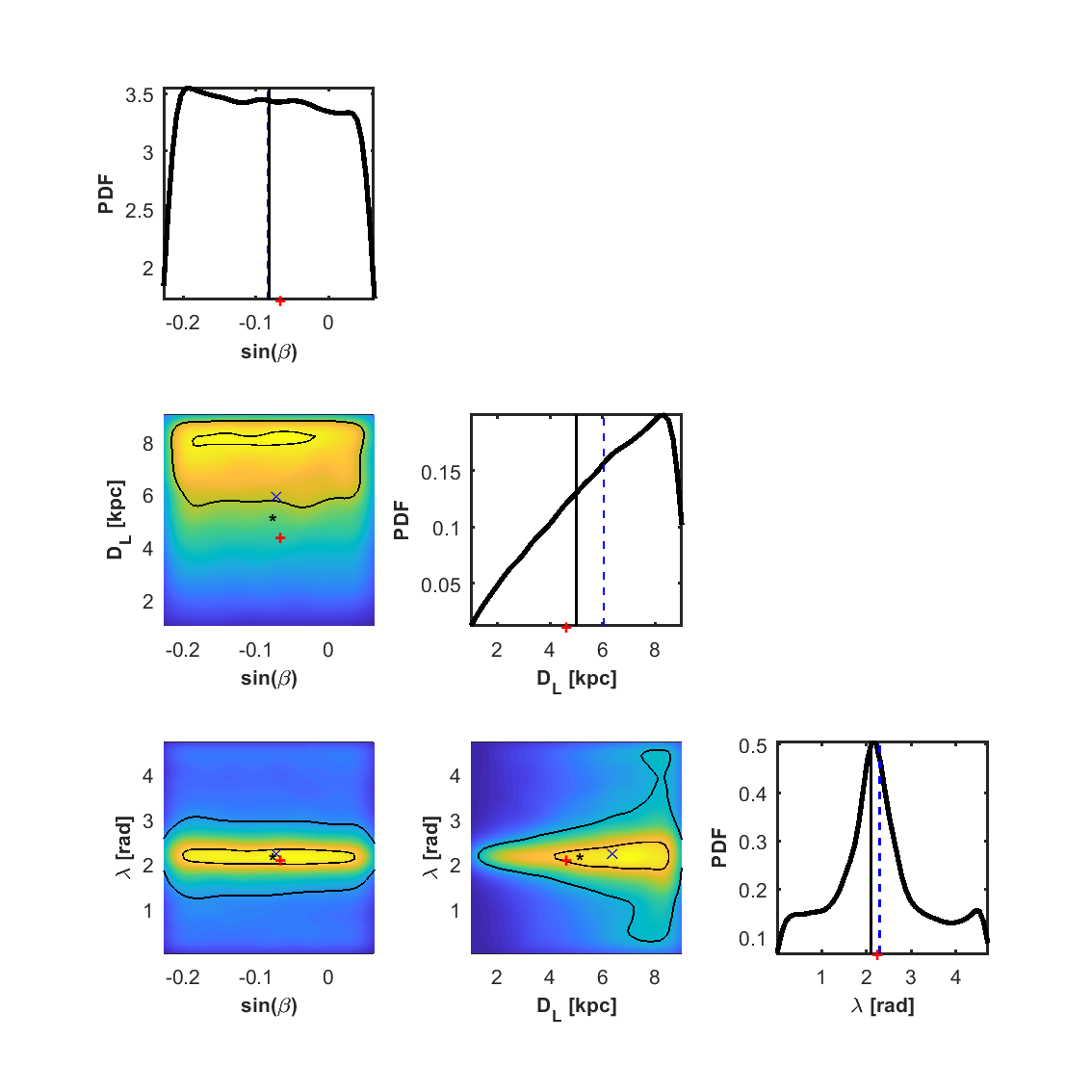}
\end{minipage}
\caption{The contract of posterior distribution of the sky position parameters  ($\lambda, sin\beta, D_L$) at best and worst orbit position of J0806 for TAIJI. The black (blue) vertical line in each one-dimensional marginalized posterior PDFs indicates the true value (statistic mean
    value). The black star (blue cross) in each two-dimensional marginalized posterior PDFs indicates the true value (statistic mean value). The red plus in each one-dimensional and two-dimensional marginalized posterior PDFs indicates best-fit value. The contours enclose the $68\%$and $95\%$ probability regions of the parameter estimation. }
\label{fig:J0806_best_worst}    
\end{figure*}

\begin{figure*}
\begin{minipage}[t]{0.49\linewidth}
    \centering
    \includegraphics[scale=0.58]{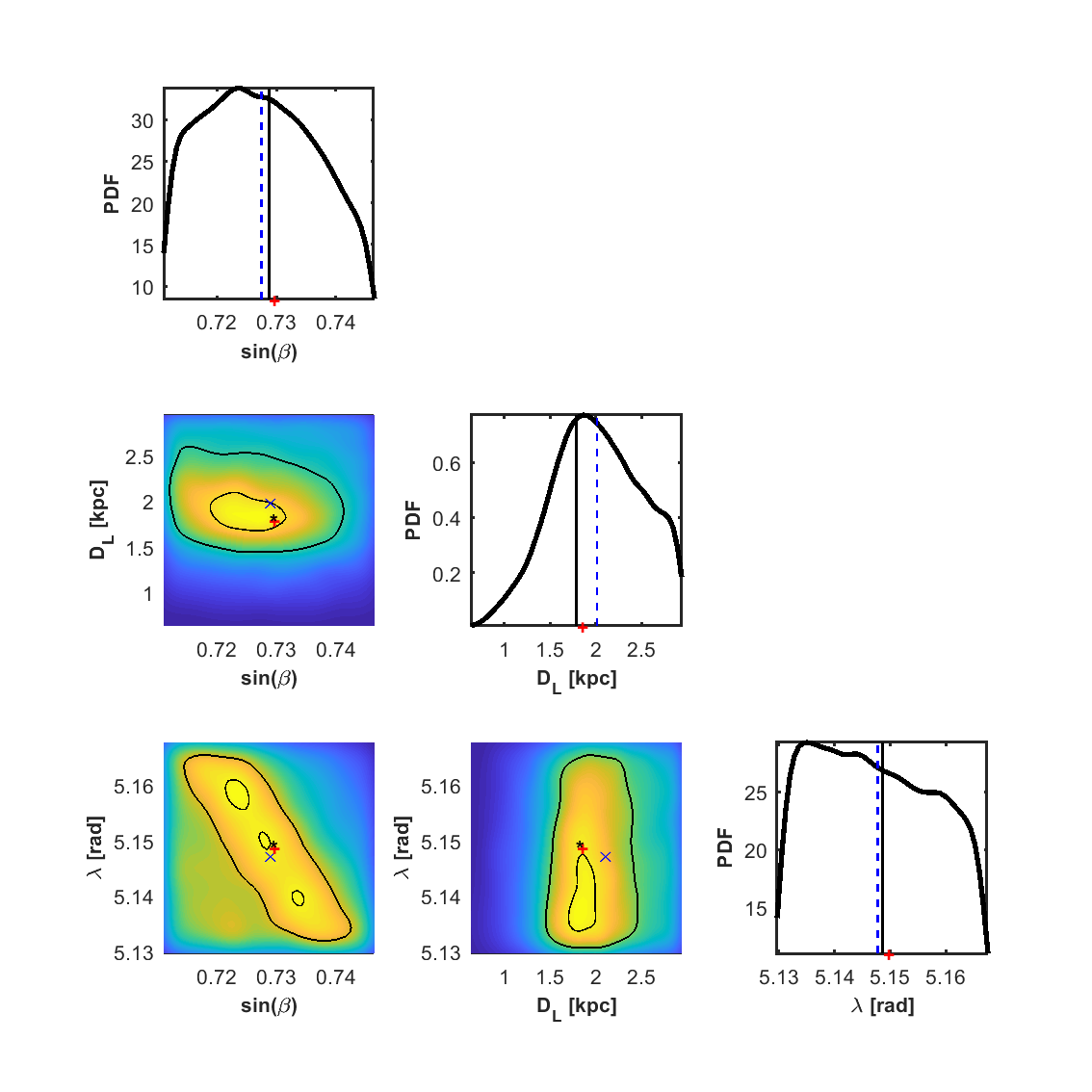}
\end{minipage}
\begin{minipage}[t]{0.49\linewidth}
    \centering
    \includegraphics[scale=0.58]{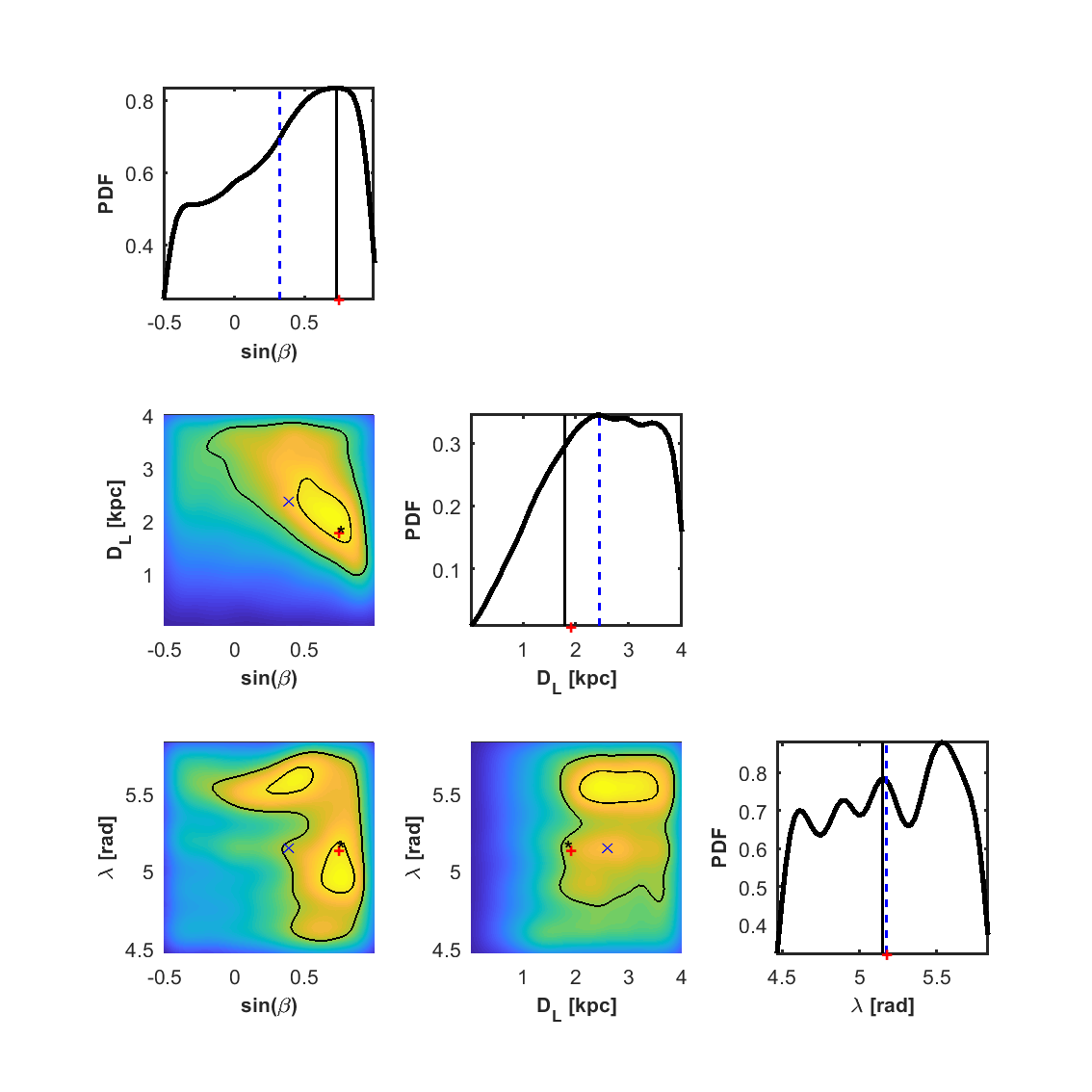}
\end{minipage}
\caption{The contract of posterior distribution of the sky position parameters  ($\lambda, sin\beta, D_L$) at best and worst orbit position of V407 Vul for TAIJI. The black (blue) vertical line in each one-dimensional marginalized posterior PDFs indicates the true value (statistic mean
    value). The black star (blue cross) in each two-dimensional marginalized posterior PDFs indicates the true value (statistic mean value). The red plus in each one-dimensional and two-dimensional marginalized posterior PDFs indicates best-fit value. The contours enclose the $68\%$and $95\%$ probability regions of the parameter estimation.}
\label{fig:v407_best_worst}    
\end{figure*}

\begin{figure*}
\begin{minipage}[t]{0.49\linewidth}
    \centering
    \includegraphics[scale=0.58]{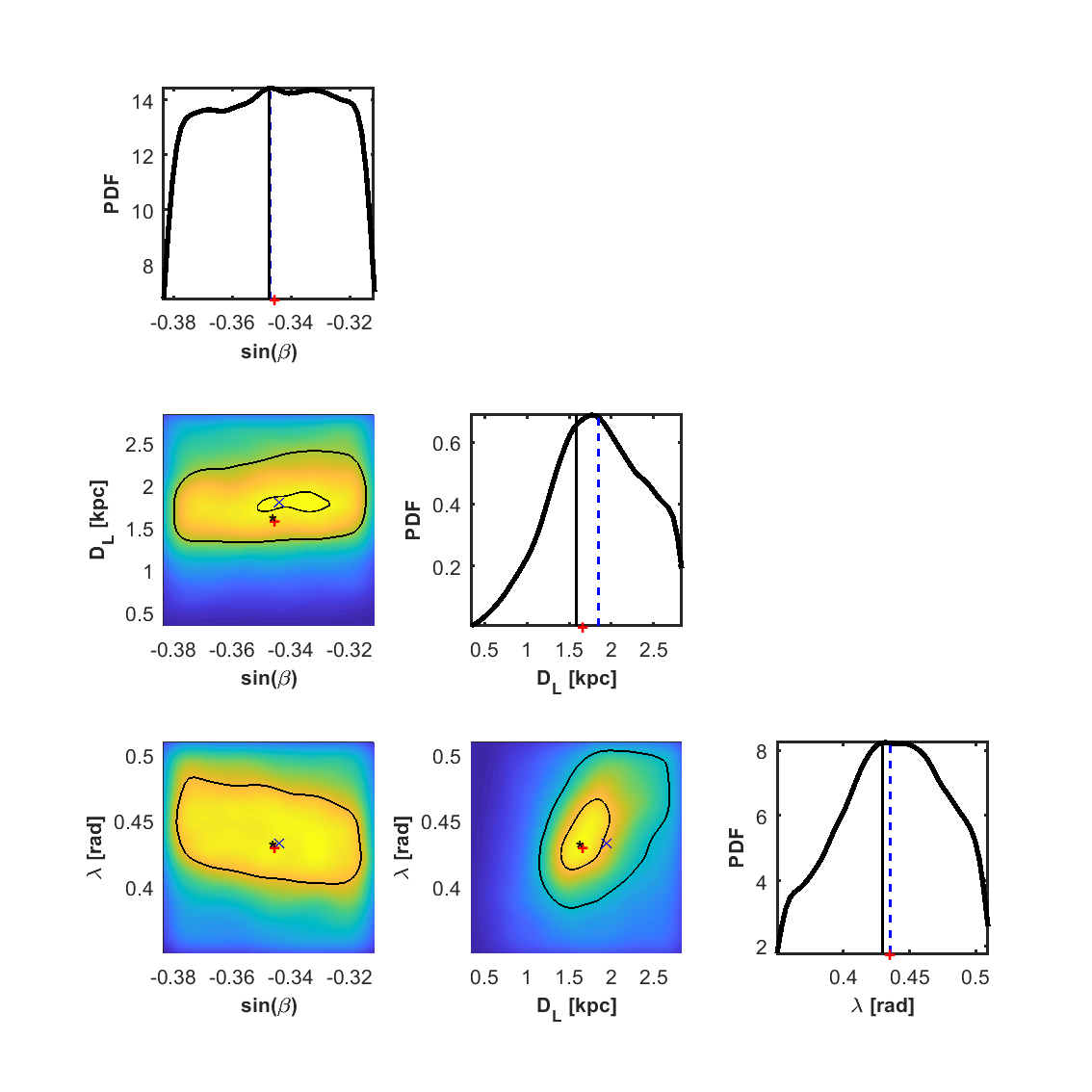}
\end{minipage}
\begin{minipage}[t]{0.49\linewidth}
    \centering
    \includegraphics[scale=0.58]{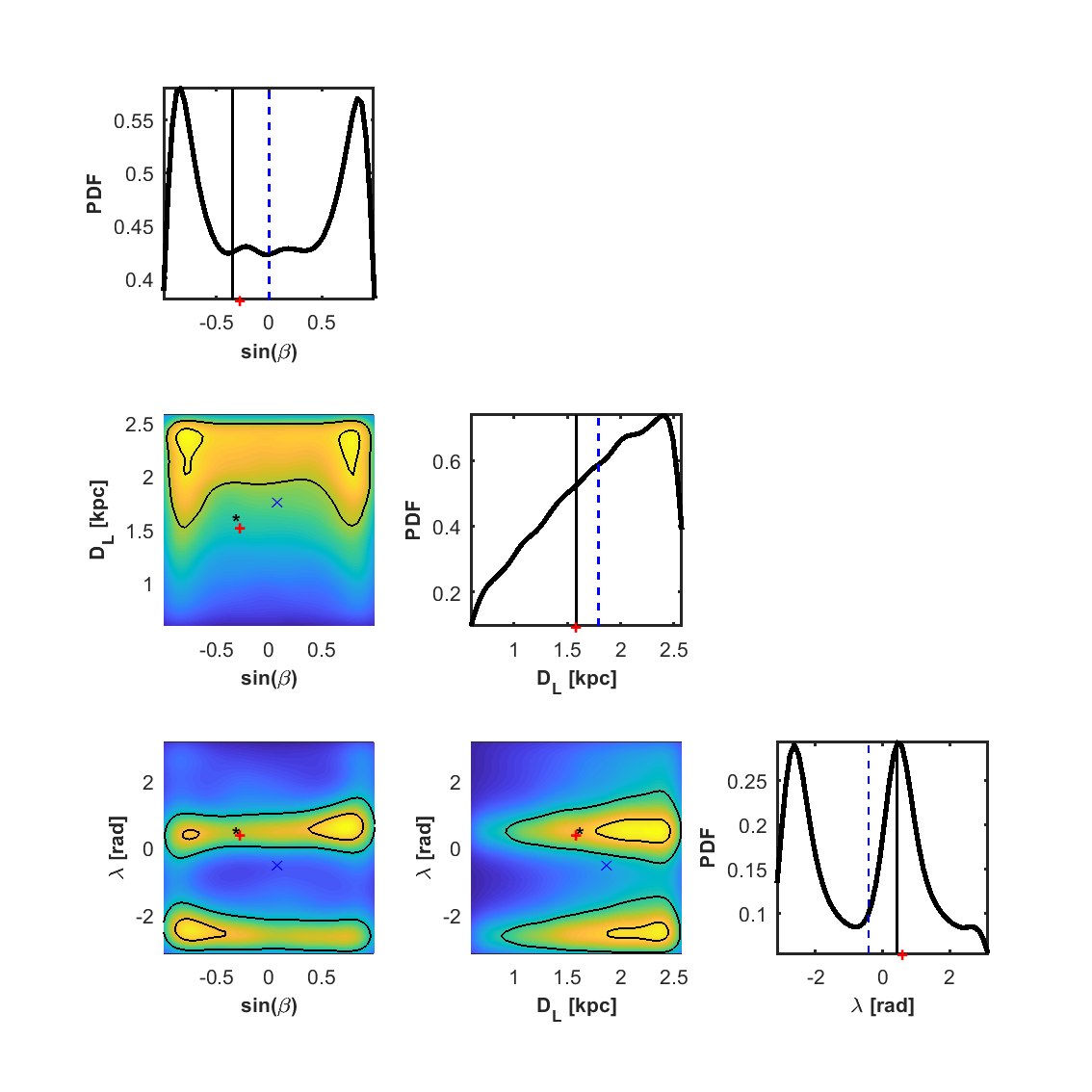}
\end{minipage}
\caption{The contract of posterior distribution of the sky position parameters  ($\lambda, sin\beta, D_L$) at best and worst orbit position of ES Cet for TAIJI. The black (blue) vertical line in each one-dimensional marginalized posterior PDFs indicates the true value (statistic mean
    value). The black star (blue cross) in each two-dimensional marginalized posterior PDFs indicates the true value (statistic mean value). The red plus in each one-dimensional and two-dimensional marginalized posterior PDFs indicates best-fit value. The contours enclose the $68\%$and $95\%$ probability regions of the parameter estimation.}
\label{fig:ES_best_worst}    
\end{figure*}

\begin{figure*}
\begin{minipage}[t]{0.49\linewidth}
    \centering
    \includegraphics[scale=0.58]{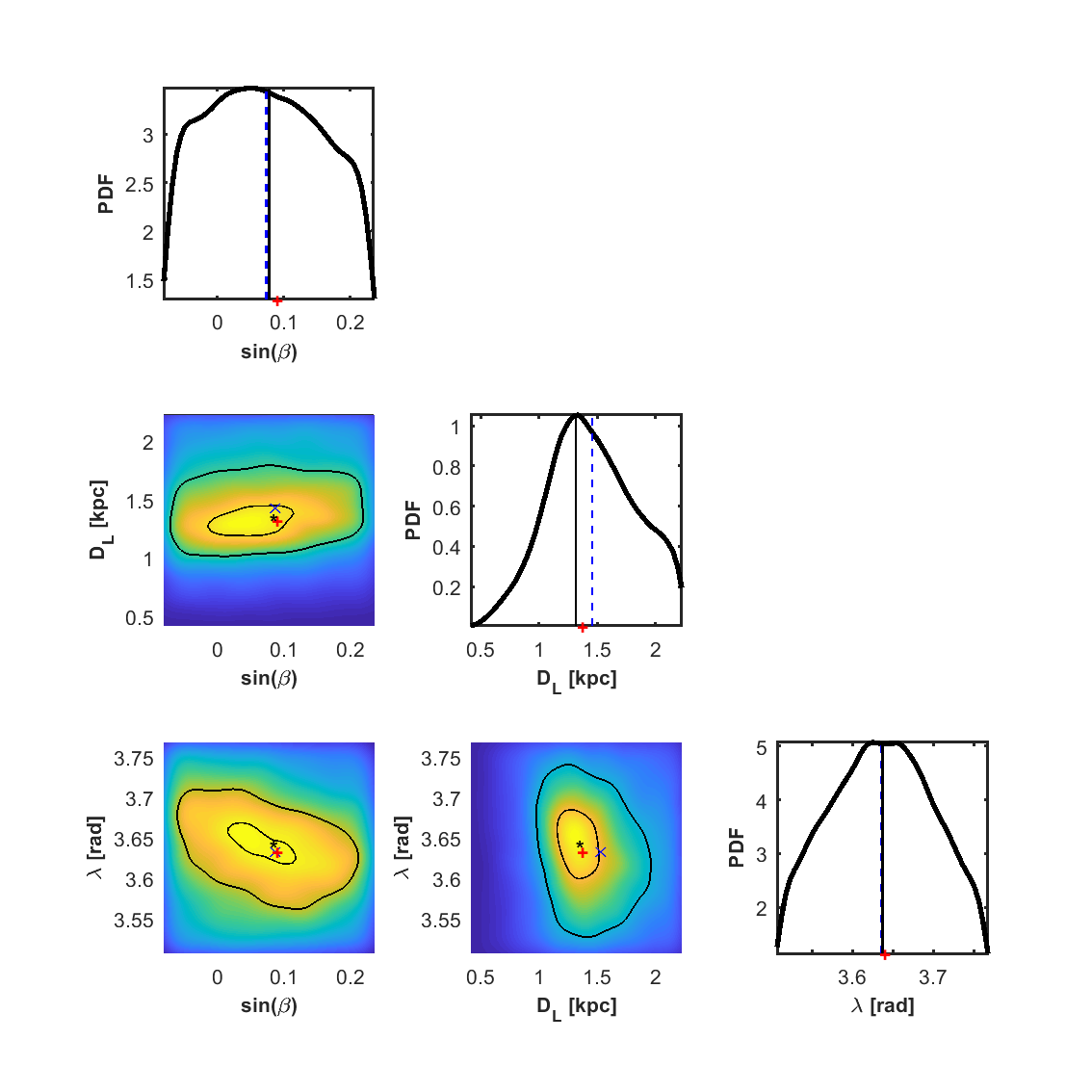}
\end{minipage}
\begin{minipage}[t]{0.49\linewidth}
    \centering
    \includegraphics[scale=0.58]{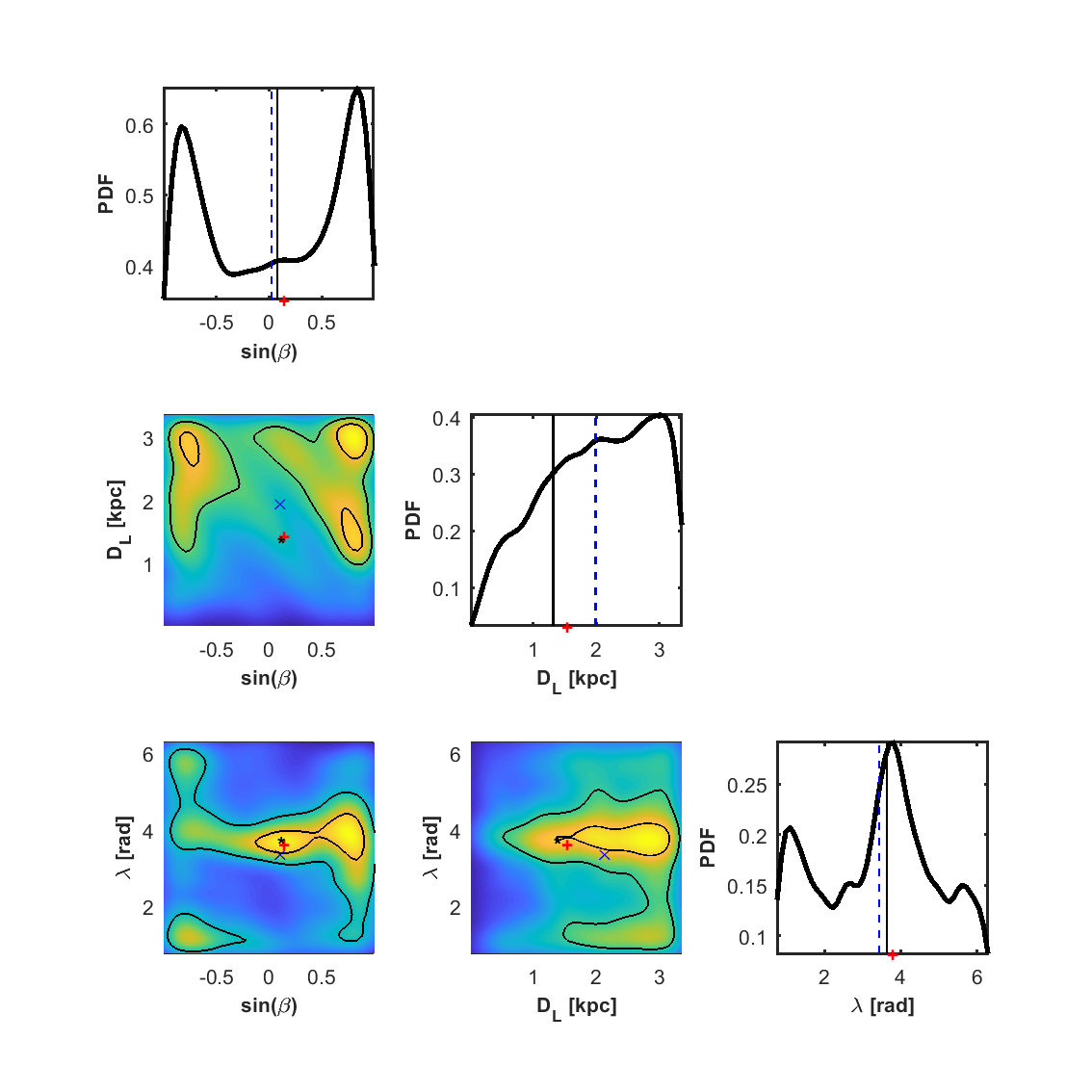}
\end{minipage}
\caption{The contract of posterior distribution of the sky position parameters  ($\lambda, sin\beta, D_L$) at best and worst orbit position of SDSSJ1351 for TAIJI. The black (blue) vertical line in each one-dimensional marginalized posterior PDFs indicates the true value (statistic mean
    value). The black star (blue cross) in each two-dimensional marginalized posterior PDFs indicates the true value (statistic mean value). The red plus in each one-dimensional and two-dimensional marginalized posterior PDFs indicates best-fit value. The contours enclose the $68\%$and $95\%$ probability regions of the parameter estimation. }
\label{fig:SDSS_best_worst}    
\end{figure*}

The contract of posterior distribution of the sky position parameters  ($\lambda, sin\beta, D_L$) at best and worst orbit position of 4 VB signals : J0806, V407 Vul, ES Cet and  SDSSJ1351 is shown in the Fig.~\ref{fig:J0806_best_worst}, Fig.~\ref{fig:v407_best_worst},Fig.~\ref{fig:ES_best_worst} and Fig.~\ref{fig:SDSS_best_worst}, respectively. We showed one-dimensional and two-dimensional marginalized posterior PDFs of the sky position parameters  ($\lambda, sin\beta, D_L$) of 4 VB signals with the best-fit value and statistic mean value. The contours enclose the $68\%$ and $95\%$ probability regions of the parameter estimation. 

According to the posterior, we find the distribution was not a standard Gaussian. So the result calculated by FIM can be not accurate. Our method gives a much more accurate result. Comparing the observations of the best and worst orbital positions simultaneously, we find that the accuracy of the parameters obtained in the best position is significantly several times higher than that in the worst position. 

\subsection{LISA-TAIJI network}

\begin{figure*}
    \centering
    \vspace{-0.35cm}
    \subfigtopskip=2pt
    \subfigbottomskip=2pt
    \subfigcapskip=-5pt
    \subfigure[VB : J0806.]{
    \includegraphics[width=0.47\textwidth]{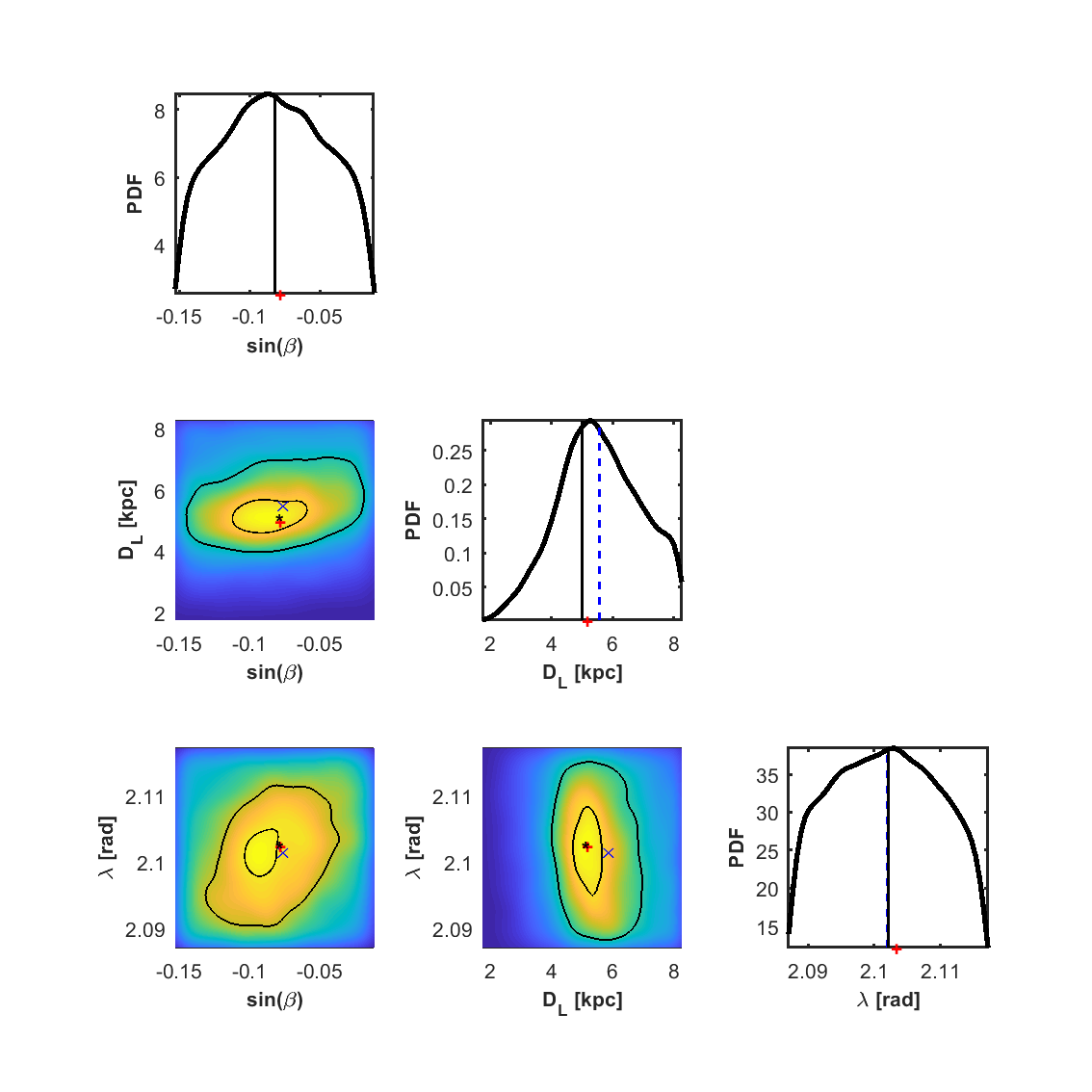}
    }
    \quad
    \subfigure[VB : V407 Vul.]{
    \includegraphics[width=0.47\textwidth]{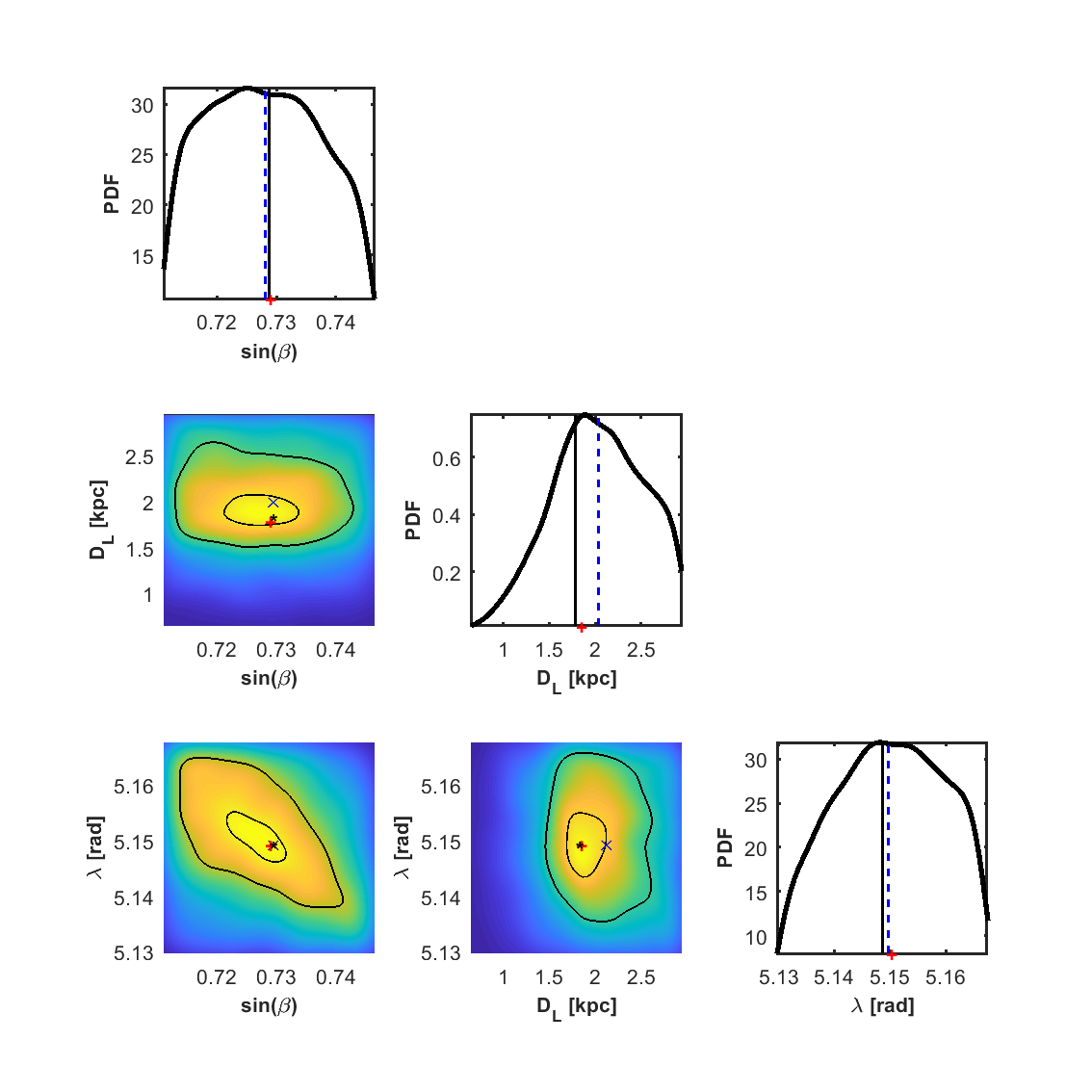}
    }
   
    \subfigure[VB : ES Cet.]{
    \includegraphics[width=0.47\textwidth]{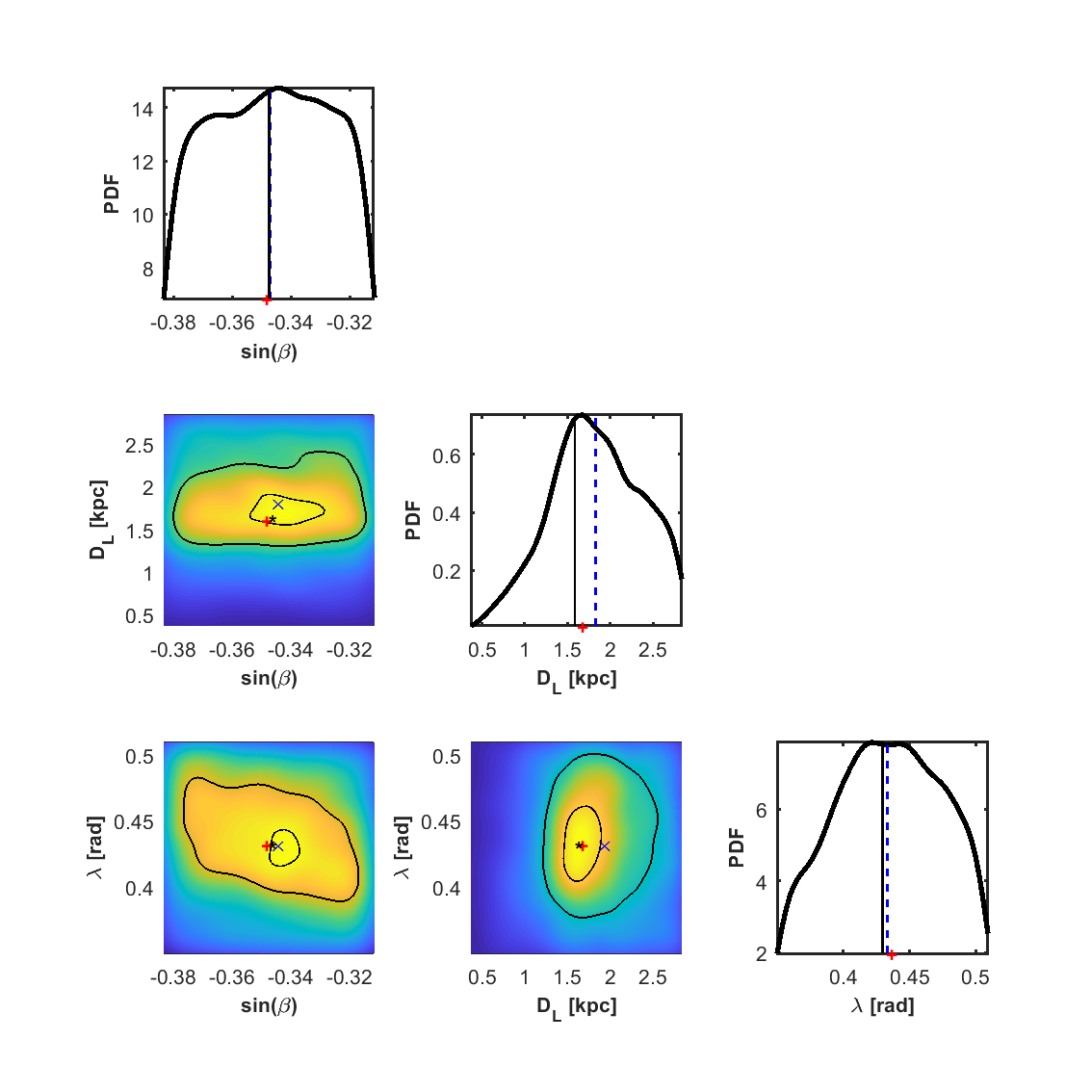}
    }
     \quad
    \subfigure[VB : SDSSJ1351.]{
    \includegraphics[width=0.47\textwidth]{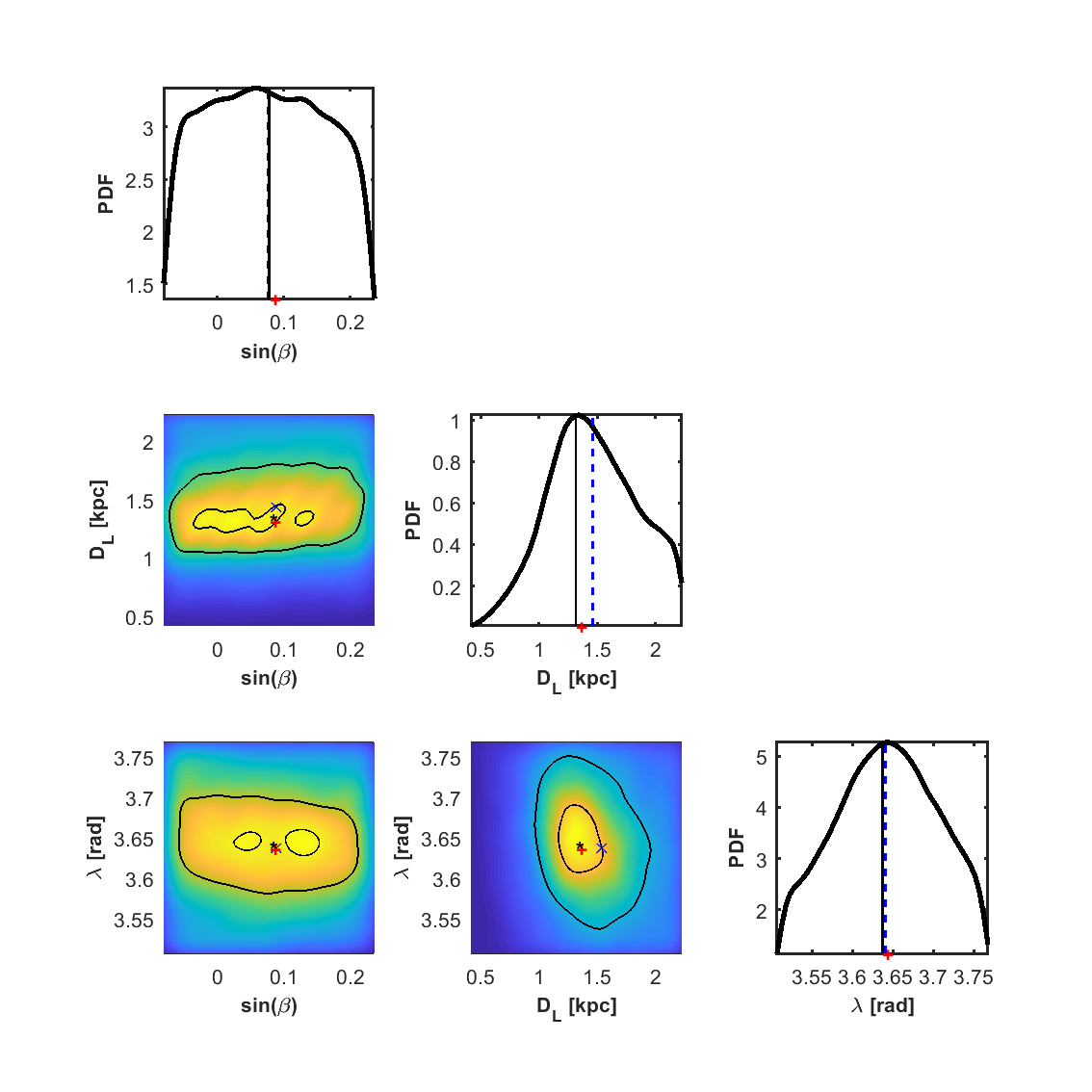}
    }
     \caption{TAIJI-LISA network: Posterior distribution of the sky position parameters  ($\lambda, sin\beta, D_L$) of 4 VB signals : J0806, V407 Vul, ES Cet, SDSSJ1351, respectively. The black (blue) vertical line in each one-dimensional marginalized posterior PDFs indicates the true value (statistic mean
    value). The black star (blue cross) in each two-dimensional marginalized posterior PDFs indicates the true value (statistic mean value). The red plus in each one-dimensional and two-dimensional marginalized posterior PDFs indicates best-fit value. The contours enclose the $68\%$and $95\%$ probability regions of the parameter estimation. }
     \label{fig:countour_4VB_network}
\end{figure*} 

Now we discuss the sky localization estimations of  TAIJI-LISA network. When we choose the arm length  $L_{TAIJI}=3 \times 10^9m$ and the orbit location variable $\kappa_{TAIJI}$ for TAIJI, we choose the arm length $L_{LISA}=2.5\times 10^9m$ and the orbit location variable $\kappa_{LISA}=\kappa_{TAIJI}-40\pi/180$. 
 We selected the best orbit position to get the posterior distribution of the sky position parameters  ($\lambda, sin\beta, D_L$) of 4 VB signals : J0806, V407 Vul, ES Cet, SDSSJ1351. 

 The posterior distribution of the sky position parameters  ($\lambda, sin\beta, D_L$) of 4 VB signals : J0806, V407 Vul, ES Cet, SDSSJ1351, respectively is shown in the Fig.~\ref{fig:countour_4VB_network}. The priors were selected as the same as the Tab.~\ref{tab:prior} at best orbit position of the TAIJI.
 We showed one-dimensional and two-dimensional marginalized posterior PDFs of the sky position parameters  ($\lambda, sin\beta, D_L$) of 4 VB signals with the best-fit value and statistic mean value. The contours enclose the $68\%$and $95\%$ probability regions of the parameter estimation.
 
By Comparison between Fig.~\ref{fig:countour_4VB_network} and the four figures above for J0806, V407 Vul, ES Cet and SDSSJ1351, the network of two detectors improve slightly the accuracy of location of the verification binaries. The reason of that result is that one GW source can not be perpendicular to both detectors of TAIJI and LISA for a short observation time. For a long observation time, the network of two detectors has a significant improvement to angular resolution.
\section{discussion}\label{sec:discussion}
As most of the known binaries are located in the Northern Hemisphere, with only a few systems located at low Galactic latitudes. It implies that the sample of current WD binaries of Galactic population may be very incomplete and biased\cite{Kupfer:2018jee,Kupfer:2023nqx}. The localizing sky position of the gravitational wave sources based on the best orbital position may cross-check the verification binaries sources obtained by astronomical observations. Additionally, this method may provide the detector's observation of significant signals in a certain sky area, which may be regarded as potential multi-messenger sources. For these sources, The targeted long-term observations may be carried out through astronomical observation methods, and more unbiased multi-messenger sources would be found.

\section{Conclusion}\label{sec:conclusion}
In this paper, we focus on localizing the sky position of gravitational wave source for the identified white dwarf binaries. 
For the first time, the effects of the periodic orbit position parameter of space-based laser interferometer detectors on the localizing the sky position of the identified double white dwarf binaries are considered. 

We present the sky maps of the WD binaries in the Milky Way before and after the projection on the detectors.
Due to the periodic orbit motion of the detectors, the results of short-time observation are related to the orbit position of the detectors.
on the best or worst orbit positions, The observations of the 4 verification binaries are performed. 
Comparing the observations of the best and worst orbital positions simultaneously, we find that the accuracy of the parameters obtained in the best position is significantly several times higher than that in the worst position.
The posterior distribution of the sky position parameters of the 4 VB sources are gotten based on TAIJI and LISA network. 
Compared with a single detector, the network of two detectors improve slightly the accuracy of location of the verification binaries. The reason of that result is that one GW source can not be perpendicular to both detectors of TAIJI and LISA for a short observation time. For a long observation time, the network of two detectors has a significant improvement to angular resolution.s

Space-based gravitational wave detection, such as LISA and TAIJI, will actually be able to observe space gravitational waves in the 2030s. This paper will provide a more rapid and accurate localization the sky position of identified sources.
\section{Acknowledgements}
We thank the anonymous referee for very helpful suggestions, which are used to improve the manuscript. This work has been supported by the Fundamental Research Funds for the Central Universities.
This work has been supported in part by the National Key Research and Development Program of China under Grant No.2020YFC2201500, the National Science Foundation of China (NSFC) under Grants Nos. 12147103, 11821505, 11975236, and 12235008, the Strategic Priority Research Program of the Chinese Academy of Sciences under Grant No. XDB23030100. In this paper, The numerical computation of this work was completed at TAIJI Cluster of University of Chinese Academy of Sciences.
\newpage
\bibliographystyle{apsrev}
\bibliography{library}
\end{document}